\documentclass[12pt]{article}
\pdfoutput=1
\usepackage{tikz}
\usetikzlibrary{matrix,arrows,positioning,shapes,snakes,fadings,decorations.pathmorphing,
decorations.pathreplacing,automata,shadows,decorations.markings,patterns}
\usepackage[thicklines]{cancel}
\usetikzlibrary{calc,shapes.callouts,shapes.arrows}
 
\usepackage{mathtools}
\usepackage{jheppub}
\usepackage{amsmath,amssymb,euscript,array,mathrsfs,appendix,ctable,marvosym,bbm,enumitem}
\usepackage{arydshln}
\usepackage{todonotes}
\usepackage{graphicx}
\usepackage{float}
\usepackage[margin=1cm,font=small]{caption}
\usepackage[margin=.2cm,font=footnotesize]{subcaption}
\usepackage[normalem]{ulem}
\usepackage{cleveref}
\usepackage[bottom]{footmisc}
\newcommand\blank[1]{#1}
\renewcommand\blank[1]{}
\def\Buildrel#1\over#2\under#3{\mathrel{\mathop{\kern0pt
#2}\limits^{#1}_{#3}}}

\usepackage{enumitem}

\def\a{\alpha}

\def\l{\lambda}

\newcommand{\Tr}{\operatorname{Tr}}

\def\B0{{\boldsymbol 0}}

\def\det{{\rm det}}

\def\ee{\boldsymbol{e}}

\def\Dbarslash{\,\,{\raise.15ex\hbox{/}\mkern-12mu {\bar D}}}
\def\Dslash{\,\,{\raise.15ex\hbox{/}\mkern-12mu D}}
\def\delslash{\,\,{\raise.15ex\hbox{/}\mkern-9mu \partial}}
\def\delbarslash{\,\,{\raise.15ex\hbox{/}\mkern-9mu {\bar\partial}}}

\def\be{\begin{equation}}
\def\ee{\end{equation}}

\DeclareFontFamily{U}{mathx}{\hyphenchar\font45}
\DeclareFontShape{U}{mathx}{m}{n}{
      <5> <6> <7> <8> <9> <10>
      <10.95> <12> <14.4> <17.28> <20.74> <24.88>
      mathx10
      }{}
\DeclareSymbolFont{mathx}{U}{mathx}{m}{n}
\DeclareFontSubstitution{U}{mathx}{m}{n}
\DeclareMathAccent{\widecheck}{0}{mathx}{"71}
\DeclareMathAccent{\wideparen}{0}{mathx}{"75}

\title{D-branes in $\lambda$-deformations}
 \author[a,b]{Sibylle Driezen,}
  \author[a,c]{Alexander Sevrin}
\author[a,b]{ and Daniel C. Thompson}
\affiliation[a]{
Theoretische Natuurkunde, Vrije Universiteit Brussel\\  \& The International Solvay Institutes\\
Pleinlaan 2, B-1050 Brussels, Belgium }
\affiliation[b]{Department of Physics, Swansea University\\Singleton Park, Swansea SA2 8PP, U.K.}
\affiliation[c]{Physics Department, Universiteit Antwerpen\\Campus Groenenborger, 2020 Antwerpen, Belgium}

\emailAdd{Sibylle.Driezen@vub.be}
\emailAdd{Alexandre.Sevrin@vub.be}
 \emailAdd{D.C.Thompson@Swansea.ac.uk}

\abstract{  We show that the geometric interpretation of D-branes in WZW models as twisted conjugacy classes persists in the $\lambda$--deformed theory.  We obtain such configurations by demanding that a monodromy matrix constructed from the Lax connection of the $\lambda$--deformed theory continues to produce conserved charges in the presence of boundaries.  In this way the D-brane configurations obtained correspond to ``integrable'' boundary configurations.  We illustrate this with examples based on $SU(2)$ and $SL(2,\mathbb{R})$, and comment on the relation of these D-branes to both non-Abelian T-duality and Poisson-Lie T-duality.  We show that the D2 supported by D0 charge in the $\lambda$--deformed theory map, under analytic continuation together with Poisson-Lie T-duality, to D3 branes in the $\eta$-deformation of the principal chiral model.  
}

\begin{document}

\pgfdeclarelayer{background layer} 
\pgfdeclarelayer{foreground layer} 
\pgfsetlayers{background layer,main,foreground layer}

\maketitle

\section{Introduction}\label{sec:lambda} 
 
The study of two-dimensional quantum field theories with boundaries has rich physical and mathematical significance.  In the context of open string theory, the boundary conditions describe D-branes, an essential non-perturbative ingredient of string theory.  More generally 2d conformal field theories with boundaries  have applications  in condensed matter systems including boundary critical behaviour, percolation models and quantum impurity problems (see e.g.\ \cite{DiFrancesco:1997nk,Cardy:2004hm}).   When the correct CFT description is available the application of boundary conformal field theory (BCFT)  can render the physics tractable \cite{Ishibashi:1988kg,Cardy:1989ir}.    Another interesting class of theories comes from integrable systems\footnote{See \cite{Babelon} or for  a recent set of useful lecture notes \cite{Zarembo:2017muf}.}; here one is lead to ask (both classically and quantum mechanically) what boundary conditions can be implemented such that the integrability of the system is preserved.    The present paper will in some way consider both contexts; we aim to study boundary effects in field theories that are integrable and at some point in the parameter space are connected to conformal field theories.  
\\
\\
Our primary motivation will be that of string theory; i.e.\ we are interested in understanding the D-branes admitted by a given curved closed string background.  In general this is a very demanding problem since one would like to have a precise CFT formulation, see e.g.\ \cite{Schomerus:2002dc}.   A simple example is provided by the Wess-Zumino-Witten  (WZW) model \cite{Witten:1983ar} describing closed strings propagating in a group manifold supported by Neveu-Schwarz flux.   Whilst BCFT can be employed here to give an algebraic description of the D-brane \cite{Kato:1996nu}  our interest will lie in the elegant geometric picture developed through a number of works \cite{Alekseev:1998mc,Felder:1999ka,Stanciu:1999id,Figueroa-OFarrill:1999cmq}. By identifying possible gluing conditions at the boundary it is determined that D-branes are described by (twisted) conjugacy classes.   For example in the $SU(2)_k$ WZW model one finds two  D0-branes and a further  $k-1$ D2-branes  that are blown up to wrap the conjugacy classes described by $S^2 \subset S^3$.
\\
\\
When an explicit CFT description is unknown and one has just a non-linear sigma-model describing strings in a curved background, giving   a precise description of D-branes is challenging.  However in a  special circumstance, namely when the sigma-model is integrable, progress can be made.  One can seek boundary conditions that preserve integrability and hope that they are amenable to a simple interpretation as D-branes.   What is meant by {\em preserve integrability}?   At a classical level it is natural to demand that the boundary condition preserves a large number of conserved charges.  There are typically two sorts of conserved charges at play, higher spin local charges and non-local charges (e.g.\ for the principal chiral model (PCM)  see \cite{Evans:1999mj} for local charges and the construction of a non-local   Yangian $\times$ Yangian structure starts with  \cite{Luscher:1977uq}).   In this work we will focus our attention on the boundary conditions that preserve some portion of the tower of non-local charges obtained from a monodromy matrix.  We anticipate that the same boundary conditions will also preserve higher spin local charges.  We shall see that this quite naturally generalises the gluing conditions used in the case of the WZW model.    This approach has its origins in  \cite{Cherednik:1985vs,Sklyanin:1988yz}  and has been used in a variety of contexts including the identification  of integrable boundary conditions  for strings in bosonic sigma models \cite{Mann:2006rh}, in Green-Schwarz sigma models  \cite{Dekel:2011ja}\footnote{In the context of the AdS/CFT correspondence see also \cite{Zoubos:2010kh} and references therein.}, for  the $O(N)$ sigma model \cite{Corrigan:1996nt,Aniceto:2017jor}, the principal chiral model \cite{Delius:2001he,Gombor:2018ppd}, open spin chains (e.g.\ \cite{deVega:1993xi,Arnaudon:2004sd} although the literature is vast) and  affine Toda field theories \cite{Bowcock:1995vp}. For methods based on the conservation of local spin charges see e.g. \cite{Ghoshal:1993tm,MacKay:2001bh,MacKay:2004rz,Moriconi:2001xz}.
\\
\\
 The present manuscript will seek to make a bridge between the two above ways of determining boundary conditions.  The $\lambda$-deformed WZW model introduced by Sfetsos in \cite{Sfetsos:2013wia} provides an ideal arena to do this.   At a classical level the $\lambda$-deformation is an integrable 1+1 dimensional field theory that depends on a parameter $\lambda \in [ 0 ,1]$.  At $\lambda = 0$ the WZW model is recovered while in a scaling limit as $\lambda \rightarrow 1$ one find the non-Abelian T-dual to the   principal chiral model.   For generic values of $\lambda$  the dynamics can be encoded by means of a Lax connection and so  a natural question to ask is what boundary conditions can be placed on a $\lambda$-model that preserve integrability.   Should one wish, once such boundary conditions are established, the limit $\lambda \to 0$ can be taken providing an alternative route to the D-brane configurations of the WZW model.  It is tricky to apply integrability techniques directly to the WZW model  due to the chiral nature of the current, so one can think of $\lambda$ as providing a convenient alleviation of this.   What we shall see in this paper is that, through a number of pleasing algebraic cancelations,  the twisted conjugacy classes define integrable boundary conditions even in the deformed model\footnote{When taken in isolation the $\lambda$-deformation is actually a marginally relevant deformation of the WZW and is not a CFT \cite{Appadu:2015nfa}.  However we continue to use the terminology of D-branes to describe boundary conditions.  It is expected  that when applied to super-cosets the $\lambda$-deformation becomes a true marginal integrable deformation \cite{Hollowood:2014qma,Appadu:2015nfa,Hollowood:2014rla} which hopefully makes this usage somewhat acceptable to a string theorist. Moreover, they provide genuine supergravity solutions when a suitable ansatz is made for the RR fluxes and the dilaton \cite{Sfetsos:2014cea,Borsato:2016ose,Demulder:2015lva,Borsato:2016zcf,Chervonyi:2016ajp}. }.  Moreover, when a quantisation condition is required on the worldvolume flux of D-branes the cancellations are such that the result is independent of the (continuous) deformation parameter $\lambda$ as it must indeed be for consistency. 
\\
A further motivation for the present study comes from duality.  To fully establish a duality one would like to have access to its action on both perturbative and non-perturbative degrees of freedom.   Consider conventional Abelian target space duality in the absence of any NS two-form potential: here the interpretation is quite simple, a T-duality transverse to a D$p$ brane   produces a D$(p+1)$ brane whilst when performed along a direction of the worldvolume  a D$(p-1)$ is produced.   This can be seen rather simply by considering the open string boundary conditions; since T-duality acts as a reflection on right-movers, Neumann $N$ and Dirichlet $D$ boundary conditions are interchanged.   For more exotic notions of target space duality, e.g.\ non-Abelian \cite{delaOssa:1992vci}  or Poisson-Lie \cite{Klimcik:1995ux},  such an understanding is less refined (although see \cite{Forste:1996hy,Borlaf:1996na,Klimcik:1995np,Klimcik:1996hp,Albertsson:2006zg,Albertsson:2007it,Hlavaty:2008jv}  and recent work in \cite{Fraser:2018atd,Cordonier-Tello:2018zdw,Borsato:2018idb}) one reason being that the geometries concerned are not flat making it harder to identify the appropriate boundary conditions.  Here however we will have access to an elegant description of D-brane boundary conditions in the $\lambda$-model whose interplay with duality can be readily studied.  For instance by taking the $\lambda \to 1$ limit on our D-brane configurations we will gain information about the brane spectrum of the non-Abelian T-dual theory.    It is also known that after appropriate analytic continuations the $\lambda$-model produces a geometry that is Poisson-Lie T-dual  to the integrable   $\eta$-deformed principal chiral model \cite{Hoare:2015gda,Sfetsos:2015nya,Klimcik:2015gba,Hoare:2017ukq}.  By examining this analytic continuation on our D-brane configurations we will too  gain information about the brane spectrum of the Poisson-Lie  T-dual theory. We study this interplay of duality with D-brane configurations in the case of the $SU(2)$ theory.  Here we find that under either the ``$\lambda \to 1$ + non-Abelian T-duality'' or ``analytic continuation + PL T-duality''  procedures, the  D2 branes discovered in the $\lambda$-model are transformed to space filling D3 branes. 
\\
\\
The structure of the paper is as follows.  We will briefly review in section \ref{sec:lambda} the saliant features of integrable $\lambda$-deformations of WZW models, in section \ref{sec:intbc} we will explain the strategy used to derive integrable boundary conditions and then apply it directly to the  $\lambda$-deformed WZW.  We illustrate this in section \ref{sec:examples} in the context of examples based on both $SU(2)$ and $SL(2,\mathbb{R})$ theories, where in the latter case we see the possibility of twisting conjugacy classes by an algebra outer automorphism.   In section \ref{sec:Tduals} we explore the relation of these D-brane configurations to both non-Abelian and Poisson-Lie T-duality. In the appendices we establish our conventions and the necessary ingredients of the general sigma model and the WZW model.

\section{$\lambda$-deformations}\label{sec:lambda} 

 First, we will briefly review the construction of \cite{Sfetsos:2013wia} in order to set up the sigma model action from which the background fields can be read off and a Lax connection  representing  the equations of motion can be found. The isotropic $\lambda$ deformation from group manifolds is obtained by starting with a sum of a PCM on $G$ for a group element $\widetilde{g} \in G$\footnote{See appendix \ref{a:conventions} for conventions.},
 \begin{equation}\label{eq:PCMaction}
 S_{\text{PCM}}(\widetilde{g}) = -  \frac{\kappa^2}{\pi} \int \mathrm{d}\sigma\mathrm{d}\tau \, \langle \widetilde{g}^{-1}\partial_+ \widetilde{g}  \widetilde{g}^{-1}\partial_- \widetilde{g} \rangle \, , 
 \end{equation}
  and a WZW model on $G$ for a different group element $g\in G$ as in eq.~\eqref{eq:WZWaction1}. Altogether, this action has a global $G_L\times G_R$ symmetry for the PCM and a $G_L(z) \times G_R(\bar{z})$ symmetry group for the WZW. One continues by gauging simultaneously the left symmetry action of the PCM and the diagonal action of the WZW acting as,
  \begin{align}
  G_L : \quad  \widetilde{g} \rightarrow h^{-1} \widetilde{g}\, , \quad
  G_{\text{diag}} :    g\rightarrow h^{-1}g h\, ,
  \end{align}
  with $h \in G$, using a common gauge field $A = A^A T_A$  transforming as,
  \begin{equation}
  A \rightarrow h^{-1} A h - h^{-1} \mathrm{d} h \ .
  \end{equation}
  The total model is made gauge invariant by replacing the derivatives in the PCM by a covariant derivative $\widetilde{D}_\pm \widetilde{g} = \partial_\pm \widetilde{g} - A_\pm \widetilde{g}$ (i.e.\ by minimal substitution),
   and by replacing the WZW model \eqref{eq:WZWaction1} with the $G/G$ gauged WZW model,
  \begin{equation}
  S_{\text{gWZW,k}}(g,A) = S_{\text{WZW},k}(g) + \frac{k}{\pi} \int \mathrm{d}\sigma\mathrm{d}\tau \, \langle A_-  \partial_+ g g^{-1} - A_+   g^{-1}\partial_- g  + A_- g A_+ g^{-1} - A_- A_+ \rangle \,    .
  \end{equation}
Finally, we can fix the gauge to $\widetilde{g} = 1$ to find,
\begin{equation}
  \begin{aligned}\label{eq:LambdaAction1}
  S_{k,\lambda}(g, A) = &\, S_{\text{WZW},k} (g) 
  - \frac{k}{\lambda \pi} \int \mathrm{d}\sigma\mathrm{d}\tau \langle A_+  O_{g^{-1}}^{-1} A_- \rangle
  \\
  &+ \frac{k}{\pi} \int \mathrm{d}\sigma\mathrm{d}\tau \, \langle A_- \partial_+ g g^{-1} - A_+ g^{-1}\partial_- g   \rangle   \,,
  \end{aligned}
  \end{equation}
%
%
%
%
where we introduced the useful operator, 
\begin{equation}
 O_g = \left(\mathbf{1} - \lambda D \right)^{-1} \, , 
\end{equation} 
given in terms of the adjoint action $D(T_A) =\mbox{ad}_gT_A= g T_A g^{-1} = T_B D^B{}_A(g)$. 
 For the isotropic $\lambda$-model, which is the model we consider throughout this paper, we have,
\begin{equation}\label{eq:IsotropicLambda}
  \lambda_{AB} = \lambda \langle T_A, T_B\rangle \equiv \lambda \eta_{AB}, \qquad \lambda = \frac{k}{k+\kappa^{2}}\,  .
  \end{equation}  
The gauge fields are now auxiliary and can be integrated out. Varying  the action $S_{k,\lambda}(g) $ with respect to $A_{\pm}$ we find the constraints,
\begin{equation}\label{eq:GaugeConstraints}
A_+ = \lambda \,O_g  \partial_+ g g^{-1}\, , \quad   
A_-  = - \lambda \,O_{g^{-1}} g^{-1}  \partial_- g\, ,
\end{equation}  
Substituting these constraints into eq.~\eqref{eq:LambdaAction1} gives,
\begin{align}\label{eq:LambdaAction2}
S_{k,\lambda}(g) &= S_{\text{WZW},k}(g) + \frac{k}{\pi}\int \mathrm{d}\sigma \mathrm{d}\tau \, R_+^A \left( \lambda^{-1} - D^T \right)^{-1}_{AB} L_-^B \, \nonumber \\
&=  \frac{k}{2\pi}\int_\Sigma  d  \sigma d\tau   L_+^A \eta_{AB} L^B_-   + \frac{k}{\pi}\int \mathrm{d}\sigma \mathrm{d}\tau \, R_+^A \left( \lambda^{-1} - D^T \right)^{-1}_{AB} L_-^B \nonumber \\  & \quad  + \frac{  k}{24\pi }       \int_{M_3}  F_{ABC} L^A \wedge L^B \wedge L^C \ ,
\end{align}
in which Lie algebra indices out of position are raised and lowered with the metric $\eta = \langle \cdot , \cdot \rangle$ and we work in term of the  Maurer-Cartan forms $L = g^{-1}dg = - i L^A T_A$  and  $R = dg g^{-1} = -  i R^A T_A$.  From the above $\lambda$ action it is straightforward to read off the target space data which can be expressed  in terms of the left-invariant forms $L^A$  as:
\begin{eqnarray}
\begin{aligned}
\mathrm{d}s^{2}_\lambda &= k \left( O_{g^{-1}} + O_g -  \eta \right)_{AB} L^A\otimes L^B  \ ,  \label{eq:MetricLamba1}\\
B_\lambda  &= B_{\text{WZW}} + \frac{k}{2} \left( O_{g^{-1}} - O_g \right)_{AB} L^A \wedge  L^B\label{eq:BfieldLambda} \ , 
\end{aligned}
\end{eqnarray} 
where we have used that $R^A = D^A{}_{B} (g) L^B$   and in which $\mathrm{d}B_{\text{WZW}} = \frac{k}{6} F_{ABC} L^A \wedge L^B \wedge L^B $.  Equivalently we can use the identity (and this will proof useful later),
\be
O_{g^{-1}} + O_g -  \eta = (1-\lambda^{2})\, O_g \,\eta\, O_{g^{-1}}=  (1-\lambda^{2}) \,O_{g^{-1}} \,\eta\, O_{g} \ ,
\ee 
to express the target space metric as,
\begin{equation}\label{eq:MetricLambda2}
\mathrm{d}s^{2}_\lambda = k\, \eta_{AB}\, E^A \otimes E^B, \qquad E=   \sqrt{1-\lambda^{2}}\, O_{g^{-1}} L  ,
\end{equation}
with $E$ the vielbein bringing us to the  flat frame. In addition the Gaussian elimination of the gauge fields, when performed in a path integral, results in a non-constant dilaton profile,
\begin{equation}\label{eq:DilatonProduced}
\Phi = \Phi_0 -\frac{1}{2} \ln \det O_{g^{-1}} \ ,
\end{equation}
in which constants are absorbed into $\Phi_0$. Finally, one can derive the classical energy momentum tensor of the $\lambda$-model to find,
\begin{equation} \label{eq:EMLambda}
T_{\pm\pm} =  k\, (1-\lambda^{2})   \langle A_\pm  \ , A_\pm\rangle\,  .
\end{equation}  
\indent There are two interesting  limits at play here \cite{Sfetsos:2013wia}; for $\lambda \rightarrow 0 $ we see from eq.~\eqref{eq:GaugeConstraints} that the fields $A_\pm$ will freeze out and, hence, one will reproduce the well-understood WZW model (see appendix \ref{a:conventions}), allowing consistency checks of analyses of the deformation. For small $\lambda$ the WZW is deformed by a current-current bilinear. In the $\lambda \rightarrow 1$ limit one reproduces the non-Abelian T-dual of the principal chiral model. This limit is more subtle and should  be taken by $k\rightarrow \infty$ in eq.~\eqref{eq:IsotropicLambda} and expanding group elements around the identity (but see section \ref{sec:Tduals} for more details).  

To establish the classical integrability of the $\lambda$ model  it is   convenient to  
 work with eq.~\eqref{eq:LambdaAction1}  but where we take the gauge fields on-shell eq.~\eqref{eq:GaugeConstraints} (although see \cite{Sfetsos:2013wia} for the proof of integrability starting from eq.~\eqref{eq:LambdaAction2}). In this way any variation of the action with respect to $A_\pm$ vanishes and calculations simplify. From  eq.~(\ref{eq:LambdaAction1}) we then find the  equation of motion for the group element $g$ to be,
\begin{equation}
D_+ \left( g^{-1} D_- g \right) = F_{+-} \ ,
\end{equation}
where we introduced the derivative $D_\pm \cdot = \partial_\pm \cdot - \left[ A_\pm , \cdot \right]$ and the field strength $F_{+-} = \partial_+ A_- - \partial_- A_+ - \left[ A_+ , A_- \right]$. 
Using the on-shell expression for $A_\pm$, eq.~\eqref{eq:GaugeConstraints}, this can be rewritten as,
\begin{equation}\label{eq:LambdaEOM}
\partial_\pm A_\mp = \pm \frac{1}{1+\lambda}\,\left[ A_+ , A_ - \right]\, .
\end{equation}
Hence, effectively we have recast one second order equation for the field $g$, by the constraints, as two first order equations for $A_\pm$. It is now straightforward to show that the following Lax connection,
\begin{equation}\label{eq:LaxLambda}
{\mathcal L}_\pm(z) = -\frac{2}{1+\lambda}\, \frac{A_\pm}{1 \mp z}\, , \qquad z \in \mathbb{C}\, ,
\end{equation}
satisfying the flatness condition $\mathrm{d} {\mathcal L} + {\mathcal L} \wedge {\mathcal L} = 0$ is equivalent to the equations of motion \eqref{eq:LambdaEOM} thereby ensuring its classical integrability  \cite{Zakharov:1973pp} (see also the next section).  

\section{Integrable boundary conditions}\label{sec:intbc}
\subsection{General methodology}\label{sec:intbcmethod}
This section closely follows  \cite{Mann:2006rh,Dekel:2011ja} for obtaining open string boundary conditions that preserve integrability based on a method first introduced by Cherdnik and Sklyanin \cite{Cherednik:1985vs,Sklyanin:1988yz} in the context of two-dimensional integrable systems.  We add here a slightly more general procedure   applicable to  integrable sigma models which has not  been clearly spelled out yet in the literature (however,  see \cite{Gombor:2018ppd} for a recent usage hereof in the case of the PCM) but which can lead to distinct integrable D-brane configurations. 

\noindent Consider first a general $(1+1)$-dimensional field theory in a spacetime (or worldsheet) $\Sigma$ parametrised by $(\tau,\sigma)$ on a periodic or infinite line with a global symmetry group $G$. The model is said to be  classically integrable when its  equations of motion can be recast in a zero curvature condition of a {\color{black} $\mathfrak{g}^\mathbb{C}$-valued} Lax connection one-form $\mathcal{L}(z)$ depending on a spectral parameter $z \in \mathbb{C}$ \cite{Zakharov:1973pp},
\begin{equation}\label{eq:LaxZeroCurv}
\mathrm{d}\mathcal{L}(z) + \mathcal{L}(z) \wedge \mathcal{L}(z) = 0 \, .
\end{equation}
The Lax $\mathcal{L}(z)$ is defined up to a local gauge transformation by a Lie group element $g(\tau,\sigma) \in G$ given by,
\begin{equation}\label{eq:LaxGaugeTransf}
\mathcal{L}(z) \;\rightarrow \; \mathcal{L}^g(z) = g \mathcal{L}(z) g^{-1} - \mathrm{d}g g^{-1}\, ,
\end{equation}
leaving the zero curvature condition \eqref{eq:LaxZeroCurv} invariant. In this case, an infinite set of conserved charges can be obtained from the usual transport matrix $T(b,a;z)$ defined by,
\begin{equation}
T(b,a; z ) = \overleftarrow{P \exp} \left(- \int^b_a \mathrm{d}\sigma\; \mathcal{L}_\sigma (\tau,\sigma ; z) \right)\, ,
\end{equation}
where we included the explicit dependence on the (worldsheet) coordinates and the arrow specifies the ordering of the integral as per \cite{Babelon}. The transport matrix satisfies the following useful properties,
\begin{align}
\delta T(b,a ;z) &= - \int^b_a \mathrm{d}\sigma T(b,\sigma;z) \delta \mathcal{L}_\sigma (\tau, \sigma ; z) T(\sigma, a;z)\, ,\\
\partial_\sigma T(\sigma,a ;z) &= - \mathcal{L}_\sigma (\tau,\sigma ;z) T(\sigma, a;z)\, ,\\
\partial_\sigma T(b,\sigma ;z) &= T(b, \sigma ;z)  \mathcal{L}_\sigma (\tau,\sigma ;z)\, ,\\
T(a,a ;z ) &=1\, ,
\end{align}
and under the gauge transformation \eqref{eq:LaxGaugeTransf} it transforms as,
\begin{equation}
T(b,a;z) \; \rightarrow \; T^g(b,a ;z) = g(\tau,  b) T(b,a ;z ) g^{-1}(\tau, a) \, .
\end{equation}
Using the flatness of the Lax $\mathcal{L}(z)$ together with the above properties, one can now show that,
\begin{equation}\label{eq:MonToTime}
\partial_\tau T(b,a ;z) = T(b,a ;z) \mathcal{L}_\tau (\tau, a;z)  - \mathcal{L}_\tau (\tau, b ;z) T(b,a;z) \, .
\end{equation}
Therefore, under periodic boundary conditions $\sigma \simeq \sigma +2\pi$ (for e.g.\ the closed string) we find that the trace of the \textit{monodromy matrix} $T(2\pi,0;z)$ is conserved\footnote{Alternatively, on the infinite line with suitable asymptotic fall-off conditions we find immediately that  $\partial_\tau T(+\infty, -\infty ; z) = 0$.},
\begin{equation}
\partial_\tau \Tr T(2\pi, 0 ;z )^n = 0 \qquad \forall \; n \in \mathbb{N} \, .
\end{equation}
Different sets of conserved charges  (local or non-local) can then be obtained from expanding the monodromy matrix or its gauge transformed form  around suitable values of the spectral parameter  leading typically to Yangian algebra's or quantum groups for the non-local sets of charges, see e.g.\ \cite{Loebbert:2016cdm}.\\
\indent For later convenience, we define here also a generalised transport matrix,
\begin{equation}
T^\Omega(b,a; z ) = \overleftarrow{P \exp} \left(- \int^b_a \mathrm{d}\sigma\; \Omega\left[ \mathcal{L}_\sigma (\tau,\sigma ; z)\right] \right)\, ,
\end{equation}
where $\Omega$ is a constant Lie algebra automorphism.
 This generalised transport matrix behaves  under time derivation as,
\begin{equation}\label{eq:GenMonToTime}
\partial_\tau T^\Omega (b,a ;z) =  T^\Omega(b,a ;z)\Omega\left[ \mathcal{L}_\tau (\tau, a;z) \right]  - \Omega\left[ \mathcal{L}_\tau (\tau, b ;z) \right] T^\Omega(b,a;z) \, , \\
\end{equation}
such that $\partial_\tau \Tr T^{\Omega}(2\pi,0;z)^{n} = 0 $ for all $n\in \mathbb{Z}$, 
and under gauge transformations as,
\begin{equation}\label{eq:GenMonGaugeTransf}
T^\Omega(b,a;z) \; \rightarrow \; \omega\left( g( b) \right) T^\Omega (b,a ; z) \omega\left(g^{-1}(a)  \right) \, ,
\end{equation}
  where the map $\omega: G \rightarrow G$ is defined as $\omega \left(e^{tX}\right) = e^{t \Omega[X]} $ for $t$ small and $X\in \mathfrak{g}$. Here we assumed the corresponding Lie group $G$ to be connected to the identity (in this case $\omega$ is a constant Lie \textit{group} automorphism also).

\noindent When the model is  considered on a finite line, e.g. $\sigma \in \left[0,\pi\right]$ in the context of sigma models  describing open strings, the charges obtained from the above procedure are generically not conserved. Similarly to the loss of conservation of momentum along the spatial direction, integrability might be spoiled. However, with the appropriate boundary conditions one can still obtain an infinite set of conserved charges from the so called \textit{boundary monodromy} object $T_b(z)$ that involves the transport matrix $T$ from the $\sigma = 0$ end of the open string to the $\sigma = \pi$ end \textit{glued} to the generalised transport  matrix $T^\Omega$ in the reflected region, i.e.\ going in the other direction. At the boundaries we include  the possibility of non-trivial effects  incorporated by the so-called reflection matrices $U_{0}$ and $U_\pi$ that are in the Lie group $G$. Hence we define\footnote{When the Lie algebra automorphism $\Omega$ is taken to be inner we recover the discussion of \cite{Mann:2006rh} up to a suitable redefinition of the reflection matrices. The possibility of $\Omega$ to be outer, however, leads to interesting non-trivial boundary conditions.}, 
 \begin{equation}\label{eq:BoundMon1}
 T_b (z) = U_0 T^\Omega_R (2\pi, \pi ; z) U^{-1}_\pi T(\pi , 0;z) \, ,  
 \end{equation}
 where the allowed boundary conditions are encoded in conditions on the reflection matrices and the automorphism $\Omega$.
%
%
%
As discussed in \cite{Mann:2006rh, Dekel:2011ja} the reflection matrices are taken to be constant in time and independent of the spectral parameter\footnote{These conditions are preferred for the interpretation of the boundary conditions as physical D-brane configurations. However  the spectral parameter independence e.g.\ might be relaxed as in  \cite{Gombor:2018ppd}   where the objective is to map the boundary conditions to known boundary scattering matrices of the quantum theory containing a free parameter.}.  The transport matrix $T_R(2\pi, \pi ; z)$ in the reflected region is constructed from the transformation $\sigma \rightarrow \sigma_R = 2\pi -\sigma$ when $\sigma\in\left[\pi,2\pi \right]$.  We will assume in the following the reflected generalised transport matrix to have the form,
 \begin{equation}\label{eq:ReflectedTransport}
 T^\Omega_R (2\pi , \pi ;z) = T^\Omega(0,\pi ; -z)\, ,  
 \end{equation}
which will indeed be the case for the $\lambda$ model.
 
In general  this strongly depends on the specific form of the Lax connection $\mathcal{L}(z)$
but the following procedure can be easily adapted to other cases. Similar to the bulk model, we impose integrability by requiring that  the time derivative of the boundary monodromy matrix is given by a commutator,
\begin{equation}\label{eq:BoundMonToTime}
\partial_\tau T_b (z) = \left[ T_b (z) , N(z) \right] \, ,
\end{equation}
for some matrix $N(z)$, such that $\Tr T_b(z)^{n} $ is conserved for any $n\in \mathbb{N}$. Explicitly we find using the formulae \eqref{eq:MonToTime} and \eqref{eq:GenMonToTime} that,
\begin{align}
\partial_\tau T_b (z) =\; &U_0 \left[ T^\Omega(0,\pi; -z) \mathcal{L}^\Omega_\tau (\pi; -z) - \mathcal{L}^\Omega_\tau (0 ;-z) T^\Omega(0,\pi ; -z) \right] U^{-1}_\pi T(\pi , 0 ; z) \nonumber \\
& + U_0 T^\Omega(0,\pi ; -z) U^{-1}_\pi \left[ T(\pi ,0 ; z)\mathcal{L}_\tau (0;z) - \mathcal{L}_\tau (\pi ; z) T(\pi,0;z) \right] \, , 
\end{align}
discarding here the $\tau$-dependence and using the notation $\mathcal{L}^\Omega(z) = \Omega[\mathcal{L}(z)]$. One can show that the integrability condition \eqref{eq:BoundMonToTime}  sufficiently holds for $N(z) =\mathcal{L}_\tau (0;z)$ when we require the following boundary conditions on both the open string ends:
\begin{equation}\label{eq:BoundCondLax1}
\mathcal{L}_\tau (\tau, 0 ;z ) = U_0 \Omega[ \mathcal{L}_\tau (\tau , 0 ; - z) ] U^{-1}_0\, ,
\end{equation}
and similarly on $\sigma = \pi$ (but where in principal the reflection matrix $U_\pi$ can be different allowing the open string to connect different D-branes). Substituting the Lax connection of the considered model  in eq.~\eqref{eq:BoundCondLax1} imposes  integrable boundary conditions on the field variables, together with consistency conditions on the automorphism $\Omega$ and the reflection matrix $U$ as will be clear in the coming subsection.\\
\indent However this is not the end of the story: the above procedure leads to sufficient conditions for   integrability of the boundary model but they are not necessary. We can cook up any exotic boundary monodromy matrix $T_b(z)$; as long as it satisfies $\partial_\tau T_b(z) = \left[ T_b (z) , N(z) \right]$ for some $N(z)$ an infinite set of conserved charges can be constructed.
It is e.g.\ an interesting possibility to consider also a gauge transformation in the reflected region ,
\begin{align}\label{eq:BoundMond2}
T_b (z , \delta) &= U_0 T^g{}_R^\Omega \left(2\pi, \pi ; \widetilde{\delta}_R ; w \right) U_\pi^{-1} T(\pi, 0; \delta ; z) \nonumber \\
&= U_0 \omega(g(0)) T^\Omega \left(0, \pi ; \widetilde{\delta} ; - w \right) \omega(g(\pi ))^{-1}U_\pi^{-1} T(\pi, 0; \delta ; z)\, ,
\end{align}
where we used eq.~\eqref{eq:GenMonGaugeTransf} and eq.~\eqref{eq:ReflectedTransport} and we included in the transport matrix\footnote{The cumbersome notation $T^g{}_R^\Omega$ represents the gauge transformed transport, acted on by the automorphism $\Omega$ and reflected.}  the possible dependence on $\delta$  representing  (multiple) deformation parameters. Moreover, in the reflected region there is the possibility that the spectral parameter and the deformation parameters change, meaning $w = F(z)$ and $\widetilde{\delta} = G(\delta)$ for some suitable functions $F$ and $G$. 
Although we will not consider this  for the $\lambda$-deformed model, we are convinced that a detailed investigation of this possibility in other integrable models will lead to interesting results and we hope to return to this point in the future.

\subsection{Applied to $\lambda$-deformations}

Having the Lax connection of the isotropic $\lambda$-deformation on group manifolds at hand one can now  derive the integrable boundary conditions corresponding to the boundary monodromy matrix $T_b (z)$ given in eq.~\eqref{eq:BoundMon1}. One can check that eq.~\eqref{eq:ReflectedTransport} indeed holds for the Lax \eqref{eq:LaxLambda}. Using now the constraints \eqref{eq:GaugeConstraints} we can write the Lax in a convenient form in terms of the Maurer-Cartan forms $L_- = g^{-1}\partial_- g $ and $R_+ = \partial_+ g g^{-1}$,
\begin{equation}
\mathcal{L}_\tau (z) =  \frac{2 \lambda}{1+\lambda} \, \frac{1}{1-z^2} \,\Big( O_{g^{-1}}[L_-] - O_g[R_+] - z \left(O_{g^{-1}}[L_-] + O_g [R_+] \right) \Big)\, ,
\end{equation}
where recall $O_g = \left( \mathbf{1} - \lambda D \right)^{-1}$. Plugging the above into the  result \eqref{eq:BoundCondLax1} and requiring the reflection matrices to be $z$-independent leads to the following boundary conditions\footnote{Here we absorbed the reflection matrices $U_{0/\pi}$ (which are essentially an additional inner automorphism action) into the definition of the automorphism $\Omega$ and discarded the indication of the open string end. However one should keep in mind that in principle $\Omega$ can be different on each end.},
\begin{align}
O_{g^{-1}}[L_-]  |_{\partial\Sigma}&= - \Omega \cdot O_g [R_+] |_{\partial\Sigma},\label{eq:LambdaGluing}
\end{align}
where, for consistency, $\Omega$ should be a constant \textit{involutive} Lie algebra automorphism.  A further restriction comes from demanding  the  (classical) conformal boundary condition (see eq.~\eqref{eq:classicalconfcond}) which requires the energy-momentum tensor to satisfy $T_{++} | = T_{--}|$.  Using the form of the stress tensor of the $\lambda$ model \eqref{eq:EMLambda}   we require that   $\Omega$  preserves the inner product $\langle \cdot , \cdot \rangle$. This is important for the boundary conditions to preserve conformal invariance in the WZW limit $\lambda\rightarrow 0$  \cite{Ishibashi:1988kg,Cardy:1989ir,Kato:1996nu,Stanciu:1999id,Alekseev:1998mc,Felder:1999ka,Figueroa-OFarrill:1999cmq}.  Summarising, $\Omega$ is a constant Lie algebra metric-preserving involutive automorphism:
\begin{equation}
\Omega \in \text{Aut} (\mathfrak{g})\,  , \quad  \Omega^{2}=1 \, ,  \quad  \Omega^T \eta \,\Omega = \eta \,.
\end{equation}
The integrable boundary conditions thus obtained reduce in the $\lambda \rightarrow 0$ limit exactly to chiral-algebra preserving symmetric D-branes of the WZW model \cite{Ishibashi:1988kg,Cardy:1989ir,Kato:1996nu,Stanciu:1999id,Alekseev:1998mc,Felder:1999ka} (in the terminology of \cite{Stanciu:1999id} the type D conditions \eqref{eq:WZWgluing}).


\subsection{Interpretation as twisted conjugacy classes}

Like in the WZW case, we desire a geometrical interpretation of the (integrable) boundary conditions \eqref{eq:LambdaGluing} as Dirichlet and Neumann  conditions on $g :\partial \Sigma \rightarrow N$ defining a D-brane $N $. In this regard it is important to realise that eq.~\eqref{eq:LambdaGluing}  takes values in the tangent space of $G$ at the identity $T_e G \simeq \mathfrak{g}$. To interpret the geometry of the D-brane configurations one needs local conditions on an arbitrary point $g\in G$ obtained by translating eq.~\eqref{eq:LambdaGluing} to $T_g G$ (which is non-trivial for non-Abelian group manifolds).  However,  as  $\lambda$ deformations of WZW theories deform only the target space data while the tangent space $T_g G$ is independent of the choice of metric, it is clear that only the size of the classical D-branes can change while their topology must be unaffected compared to the well-known WZW branes. We will show here  that this is indeed the case leading to   (integrable) boundary conditions  corresponding to D-brane configurations that are (twisted) conjugacy classes \cite{Stanciu:1999id,Felder:1999ka,Alekseev:1998mc}.

First, we should split the tangent space $T_g G$ at $g\in G$ orthogonally to the D-brane  $N$ with respect to the $\lambda$-deformed metric \eqref{eq:MetricLambda2} (assuming the metric restricts non-degenerately to $N$),
\begin{equation}
T_g G = T_g N \oplus T_g N^\perp \ .
\end{equation}
Important here  is that the object $\widetilde{\Omega}_g \equiv O^{-1}_{g^{-1}}\cdot \Omega \cdot O_g$ gluing left to right currents in  eq.~\eqref{eq:LambdaGluing} is easily shown to preserve the deformed metric  \eqref{eq:MetricLambda2} at $g$ provided that $\Omega$ preserves the inner product $\eta = \langle \cdot , \cdot \rangle$. Indeed, writing the metric as 
\begin{equation}
G_\lambda = k E^T \eta E = k\,(1- \lambda ^2)  L^T O_{g^{-1}} \,\eta \,O_g L = k\,(1- \lambda ^2)  L^T O_g \,\eta \,O_{g^{-1}}L\,,
\end{equation}
 we have $\widetilde{\Omega}^T_g G_\lambda \widetilde{\Omega}_g = G_\lambda$.
%
 %
 Moreover, it is straightforward to show that the deformed metric at $g$ is  invariant under an adjoint action by $g$, i.e. $D^T G_\lambda D = G_\lambda$.\\
 \indent With these properties of the deformed metric at hand we can now exactly follow the  procedure used in the WZW case of \cite{Stanciu:1999id,Figueroa-OFarrill:1999cmq} for finding the tangent space $T_g N$ to the D-brane when treating $\widetilde{\Omega}_g $ as the gluing matrix. This leads to,
 \begin{equation}
 T_g N = \left\{ u - O^{-1}_{g^{-1}} \cdot\Omega  \cdot O_g [g^{-1} u ]g \; | \; \forall u \in T_g G \right\}.
 \end{equation}
 Using the transitivity property of left and right translations on group manifolds, together with $\Omega$ being an automorphism (and thus bijective), there exists for every $u \in T_g G$ a Lie algebra element $X$ such that, 
 \begin{equation}
  u = - g O^{-1}_{g^{-1}} \Omega [X]\,.
 \end{equation}
 Hence, $T_g N$ is equivalently given by,
\begin{align}
T_g N &= \left\{ O_g^{-1}\left[ X \right] g - g O^{-1}_{g^{-1}}\Omega \left[ X \right]\; | \;\forall X \in \mathfrak{g}\right\} \nonumber\\
&= \left\{ \left( \mathbf{1} + \lambda \Omega \right)[X] g - g \left( \lambda\mathbf{1} +  \Omega \right) [X] \;| \; \forall X \in \mathfrak{g}\right\} \, .
\end{align}
Contrary to the WZW case, we have here the extra property on the Lie algebra automorphism $\Omega$ that it is involutive, $\Omega^2 =1$, and thus defines a symmetric space decomposition of the Lie algebra $\mathfrak{g}$:
\begin{equation}
\mathfrak{g} = \mathfrak{h} \oplus \mathfrak{k}\, ,
\end{equation}
where $\Omega[\mathfrak{h}] = \mathfrak{h}$\,,  $\Omega[\mathfrak{k}] = -\mathfrak{k}$ and,
\begin{equation}
\left[\mathfrak{h}, \mathfrak{h} \right] \subset \mathfrak{h}\,, \qquad \left[\mathfrak{h} , \mathfrak{k} \right] \subset \mathfrak{k}\,, \qquad \left[\mathfrak{k} , \mathfrak{k} \right] \subset \mathfrak{h}\,  .
\end{equation}
Hence, also $T_g N$ splits accordingly,
\begin{align}
T_g N = \left\{ (1+\lambda) \left(H g - g H \right)\; |\; \forall H \in \mathfrak{h}\right\} \oplus \left\{ (1- \lambda) \left( K g + g K\right)  \; |\; \forall K \in \mathfrak{k} \right\} , 
\end{align}
It is now possible to rescale $(1+\lambda)H \rightarrow H$ and $(1-\lambda)K \rightarrow K$ such that,
\begin{align}
T_g N &= \left\{ H g - g H\; |\; \forall H \in \mathfrak{h}\right\} \oplus \left\{K g + g K\; |\; \forall K \in \mathfrak{k} \right\} \nonumber \\
&= \left\{ X g - g \Omega [X] \; | \;\forall X \in \mathfrak{g} \right\}.
\end{align}
As expected, this is exactly the tangent space to the twisted conjugacy class $C_\omega (g)$ defined by\footnote{Recall the definition of the map $\omega : G \rightarrow G$ under eq.\ eq.~\eqref{eq:GenMonGaugeTransf}.},
\begin{equation}\label{eq:TwistConjClass}
C_\omega (g) = \left\{ h\, g\, \omega (h^{-1}) | \; \forall h \in G \right\},
\end{equation}
shown explicitly in  \cite{Stanciu:1999id,Figueroa-OFarrill:1999cmq}. Hence, the worldvolumes $N$ of the integrable D-brane configurations lie on twisted conjugacy classes $ C_\omega (g)$  of the group $G$ on which the deformation is based.  In the $\lambda$-deformed background only the size of the branes (determined by the induced deformed metric)  change, as also illustrated in the following section. The twisted conjugacy classes are classified by the quotient $ \text{Out}_0 (G) = \text{Aut}_0 (G) / \text{Inn}_0 (G)$    of metric-preserving outer automorphisms of $G$ since two automorphisms that are related by an inner automorphism in $\text{Inn}_0 (G)$ lead to twisted conjugacy classes that differ simply by a group translation  \cite{Figueroa-OFarrill:1999cmq}. However, the involution condition $\omega^2 =1$ from integrability does not allow these group translation to be arbitrary and in practice there will only be a small number of them, depending on the dimensionality of $G$. When the group automorphism $\omega$ is taken to be the identity element of $ \text{Out}_0 (G)$, the twisted conjugacy classes reduce to regular conjugacy classes \cite{Alekseev:1998mc}.

Note finally that, using Frobenius' integrability theorem and $\Omega$ being an automorphism,  the D-brane $N$ is a submanifold of $G$ eliminating the possibility of intersecting integrable D-brane configurations \cite{Stanciu:1999id,Figueroa-OFarrill:1999cmq} using the methodology outlined in  section \ref{sec:intbcmethod}.

\section{Examples}\label{sec:examples} 

In this section we will apply the above observations explicitly in two simple examples: the $\lambda$ deformation based on the $SU(2)$ and $SL(2,\mathbb{R})$ Lie groups. In the compact $SU(2)$ case only regular conjugacy classes exist and we will derive the boundary equations explicitly for them (necessary for section \ref{sec:Tduals}). Moreover we will show that the flux quantisation condition  remains independent of $\lambda$. We will perform a semi-classical analysis of the spectrum of quadratic fluctuations  and demonstrate that turning on $\lambda$ will lift the zero modes of the WZW branes. In the non-compact $SL(2,R)$ the study of  both regular and twisted conjugacy classes is possible. Here we will focus  on the classical aspects showing only the twisted conjugacy classes to be (classically) physical in contrary to the regular conjugacy classes. Moreover we   find  again the flux quantisation to remain independent of $\lambda$. 

\subsection{The $S^3$ deformation}\label{sec:su2}
For $\mathfrak{su}(2)$ the metric-preserving algebra automorphisms form the group of rotations $SO(3)$ while  $Out_0 (SU(2))$ is known to be trivial. The  D-branes in the $S^3$ manifold therefore lie on regular conjugacy classes. In light of this we choose  the following convenient parametrisation for the $SU(2)$ group element, 
\be\label{eq:su2para}
\left(
\begin{array}{cc}
 \cos \alpha +i \cos \beta\,  \sin \alpha  & e^{-i \gamma } \sin \alpha\,  \sin
   \beta  \\
 -e^{i \gamma } \sin \alpha  \,\sin \beta  & \cos \alpha -i \cos \beta\,  \sin
   \alpha  \\
\end{array}
\right)\,,
\ee 
such that the   $S^3$ is given by an $S^2$ parametrised  by $\beta \in [0,\pi]$ and $\gamma\in[0,2\pi]$ fibred over an interval  $\alpha \in [0 , \pi]$.  Regular conjugacy classes  are distinguished by $\Tr(g) = 2 \cos \alpha $ constant and so correspond to this $S^2$. Note that integrability  only  allows group translations from involutive inner automorphisms that correspond here to rotations over an angle $\pi$. For the WZW analysis of these D2-branes see  we refer to  \cite{Bachas:2000ik,Pawelczyk:2000ah,Stanciu:1999nx,Figueroa-OFarrill:2000gfl}.  

We first recall the target space geometry of the $\lambda$-deformed theory \cite{Sfetsos:2013wia},
\begin{equation}
\begin{aligned}
ds_\lambda^2 &= 2k \left( \frac{1+\l}{1-\l}\, d\alpha^2 + \frac{1- \lambda^2}{\Delta} \,\sin^2 \alpha ( d\beta^2 +\sin^2 \beta d\gamma^2)  \right) \ , \\
H_\lambda  &= 4 k\, \frac{ 2 \lambda \Delta + (1-\lambda^2 )^2 }{\Delta^2 } \,\sin^2\alpha \sin \beta d\alpha \wedge d\beta \wedge d\gamma \,,\\
e^{-2 \Phi } &= e^{-2 \Phi_0 } \Delta \ , \quad   
\end{aligned}
\end{equation}
where $\Delta = 1+ \lambda^{2} - 2\lambda \cos 2\alpha$. Here we note that in performing the Gaussian integration to arrive at the $\lambda$ model a dilaton \eqref{eq:DilatonProduced} is produced. 
In the metric observe that, and this will be important, that the deformation leaves the $S^2$ intact changing only the radius of this sphere as it is fibred over $\alpha$.  

The integrable boundary condition obtained from \eqref{eq:LambdaGluing} with $\Omega=1$ reads,
\be\label{eq:su2bc}
\begin{aligned}
\partial_- \alpha &= - \partial_+ \alpha \,,\\ 
(1+\lambda^2 -2 \lambda \cos 2\alpha ) \partial_-\beta &=  \left( 2 \lambda - (1+\lambda^2) \cos 2\alpha \right) \partial_+ \beta  - (1-\lambda^2  )  \sin \beta \sin 2\alpha \partial_+ \gamma\,, \\ 
 (1+\lambda^2 -2 \lambda \cos 2\alpha ) \partial_-\gamma &= \left(2 \lambda - (1+\lambda^2) \cos 2\alpha \right) \partial_+ \gamma + (1-\lambda^2  ) \csc \beta \sin 2\a \partial_+ \beta\,  .
\end{aligned} 
\ee 
It is immediately clear that $\alpha$ obeys a Dirichlet boundary condition and $X^a = \{ \beta  , \gamma\}$ obey (generalised) Neumann boundary conditions \eqref{eq:neumann} which should take the standard form,
\be
\widehat{G}_{ab}(X) \partial_\sigma X^b  = {\cal F}_{ab} \partial_\tau X^b\,  , 
\ee
in which ${\cal F}_{ab} = \widehat{B}_{ab} + 4\pi  F_{ab}$\footnote{For the level $k$ to obey the conventional quantisation we have used units in which $\alpha^\prime =2$.} (see also appendix \ref{a:sigmamodel}).  We first re-express the Neumann boundary condition as,
\be 
\partial_\sigma \beta = - \frac{\lambda -1}{\lambda +1} \,\cot \alpha \sin \beta \partial_\tau \gamma \,, \, \quad 
\partial_\sigma \gamma =   \frac{\lambda -1}{\lambda +1}\, \cot \alpha \csc \beta \partial_\tau \beta\, , 
\ee
and making use of the metric restricted to   $\alpha = \text{const}$ we extract the two-form, 
 \be
{\cal F} = \frac{k}{2 \Delta} (1-\lambda)^2 \sin 2 \alpha \sin\beta d\beta \wedge d \gamma  \,.
\ee 

We can now evaluate the DBI action \eqref{eq:DBIaction}, 
\be
S_{\text{DBI}} =T_2  \int e^{-\Phi} \sqrt{ \widehat{G}+{\cal F}} =  4 \pi  k\, T_2 \,e^{-\Phi_0}  \sin \alpha \,.
\ee 
where we absorbed a factor of $1-\lambda$ into the constant dilaton $e^{-\Phi_0}$.
Naively this would suggest that the only stable D-branes are the ones for which this quantity is minimised i.e. the   D0 branes located at $\alpha = 0$ and $\alpha = \pi$. However the flux quantisation stabilises the branes in other locations in a slightly subtle way.    There are two well defined forms at play; the  NS three form $H$ and a two-form on the D-brane submanifold $\omega$ which is, by virtue of the construction, equal to the two-form ${\cal F}$.   The quantisation is now a statement that the relative cohomology class $[(H, \omega )]$ be integral \cite{Klimcik:1996hp,Gawedzki:1999bq,Figueroa-OFarrill:2000lcd,Stanciu:2000fz} demanded by consistency for the definition of the WZ term  in the action \eqref{eq:LambdaAction1} (see also appendix \ref{a:conventions}).   Put plainly,  the difference in periods of $\omega$ over the D2-brane worldvolume  $N$ and $H$ on an extension   $B$  whose boundary  is the D2-brane $  \partial B = N $ should be integral i.e. 
\be\label{eq:fluxquant}
\frac{1}{4\pi}\int_{N} {\cal \omega} - \frac{1}{4\pi}  \int_{B} H  \; \in \; 2\pi\, \mathbb{Z} \ .
\ee
For the case at hand, let us locate the D2 brane at $\alpha = \alpha_\star$ and  integrate $H$ over an extension $\alpha \in [0, \alpha_\star]$.  This yields 
\be
\int_{B} H  = - 2  k \alpha_\star + \frac{ k (1-\lambda )^2 \sin  2\alpha_\star }{
\Delta_\star } \ . 
\ee
 At the same time we have 
\be
\int_{N} \omega   =   \frac{ k (1-\lambda )^2 \sin  2\alpha_\star }{
\Delta_\star } \ ,
\ee
so that in the combination entering the quantisation condition  eq.~\eqref{eq:fluxquant} all dependence on the deformation parameter $\lambda$ drops out and one recovers the conventional result for the WZW model; there are, in addition to the D0-branes,   stabilised D2's located at\footnote{We are assuming throughout that we are in the semi-classical regime and so  ignore any consequences of the shift $k\to k+2$ \cite{DiFrancesco:1997nk}.  } 
\be
 \alpha_\star  =  \frac{n \pi }{k}  \ , \quad  n= 1 \dots k-1 .
\ee

Let us now evaluate the dynamics of fluctuations of this D-brane.  For that we need to consider dependence on the target space time coordinate; i.e. we need to introduce an extra time-like dimension to the target space.   We will choose a synchronous gauge and let the coordinates of the worldvolume of the D2 be $X^a = (t, \beta, \gamma)$.  Following \cite{Bachas:2000ik} we examine fluctuations of the transverse scalar and worldvolume gauge field in  the DBI action,
\be
S= T_2 \int e^{-\Phi} \sqrt{-  \det ( \widehat{G}_{ab} + {\cal F}_{ab} )  } ,
\ee
where ,
 \be 
 {\cal F}_{ab} = B_{ab} + 2 \pi \alpha^\prime (\partial_a A_b - \partial_b A_a) \ . 
 \ee
 For direct comparison to  \cite{Bachas:2000ik}  we reinstate explicit factors of $\alpha^\prime$.  It should be emphasised that the metric $\widehat{G}$ needs to be pulled back to the worldvolume according to,
 \be
\widehat{G}_{ab} = G_{\mu \nu} \partial_a X^\mu \partial_b X^\nu  \ , 
 \ee
which induces derivatives for fluctuations in the transverse scalar.    To perform this analysis it is helpful to pick a gauge for the antisymmetric field \cite{Sfetsos:2013wia},
\be
B = - k \alpha^\prime \left( \alpha - \frac{ (1- \lambda)^2}{\Delta} \,\cos\alpha \sin\alpha  \right) \sin \beta d\beta \wedge d\gamma \ . 
\ee 
Then we  make the ansatz for fluctuations, 
\be
\alpha = \alpha_\star + \delta(X) \ , \quad A_t =0 \ , \quad A_\beta = \frac{k}{2\pi} a_\beta(X) \ , \quad A_\gamma = \frac{k}{2\pi} \alpha_\star (\cos \beta -1 ) + \frac{k}{2\pi} a_\gamma(X)  .
\ee  
Now the procedure is to expand the DBI action to quadratic order in the fluctuation and to extract classical equations of motion\footnote{As in \cite{Bachas:2000ik} in the expansion there is a linear term proportional to the fluctuation of the quantised D0 charge which must necessarily vanish and so we neglect it in what follows.}.  The intermediate steps of this calculation are extremely unedifying and made algebraically complicated by the appearance of the function $\Delta(\alpha)$ in various places.  However, somewhat remarkably to a large extent all of the complications cancel out to leave a very simple result.    In terms of the covariant fluctuation $g(t, \beta , \gamma)  =  -\frac{1}{\sin \beta} \left( \partial_\beta \alpha_\gamma -\partial_\gamma \alpha_\beta \right)$  we find equations of motion,
 \be
 \frac{d^2}{dt^2} \left(\begin{array}{c} \delta \\ g \end{array} \right) =  -  \frac{ 1}{k \alpha^\prime}    \frac{1+\lambda^2 }{1-\lambda^2}  \left(\begin{array}{c c} 2 + \frac{(1+\lambda)^2}{1+\lambda^2}  \Box & 2 \\  2 \, \Box  & \frac{(1+\lambda)^2}{1+\lambda^2}  \Box\end{array}\right)  \left(\begin{array}{c} \delta \\ g \end{array} \right)  \ ,
 \ee
 in which $\Box$ is the Laplacian on the $S^2$.  This operator can be diagonalised in terms of an expansion in spherical harmonics.  In the $l^{th}$ sector (i.e. where $\Box = l ( l+1)$) we find that the eigenvalues are,  
 \be
 \frac{1}{k (1-\lambda^2)} (1+l) \left( (1+\lambda)^2  l  +2 (1+\lambda^2) \right) \ , \quad  \frac{1}{k (1-\lambda^2)} l \left( l (1+\lambda)^2 - (1- \lambda)^2 \right) .
 \ee 
These are all positive, hence stable fluctuations of positive mass.  Since $g$ carries no $s$-wave as a consequence of flux-quantisation and  that the s-wave of $\delta$ has a frequency squared of   $\frac{ 2}{k \alpha} \frac{1+\lambda^2 }{1-\lambda^2} $ -- it is not a moduli.   It is interesting to notice that the $p$-wave triplet (i.e. $l=1$) acquires a positive mass for $\lambda  \neq 0$; this lifting of zero modes is a reflection of the fact that  one of the $SU(2)$ symmetries is broken in the target space metric and so we are not longer free to move the $S^2$ of the D-brane about the $S^3$.  Put another way, in the undeformed theory these zero modes on the worldvolume are associated Goldstone modes from  breaking the $SU(2)$ global symmetry of the target space by the D-brane; in the deformed theory there is no longer such an $SU(2)$ symmetry to be broken and hence no corresponding Goldstone.   A further feature of the spectrum is that it inherits a $\mathbb{Z}_2$ invariance $\lambda \to \lambda^{-1}$, $k \to - k$ displayed by the $\lambda$-deformed worldsheet theory \cite{Itsios:2014lca}. 
 
\subsection{The AdS$_3$ deformation}

\noindent For the $\mathfrak{sl}(2,\mathbb{R})$ algebra it can be shown that the metric-preserving automorphisms form the group $SO(1,2) = SO(1,2)^+ \cup SO(1,2)^-$ where  $SO(1,2)^+$ correspond to the usual rotations and boosts, while $SO(1,2)^-$ transformations are obtained from an additional reflection and time-reversal. The metric-preserving outer automorphism group  $\text{Out}_0(SL(2,\mathbb{R}))$ descends from the latter and can be shown to have, besides the identity,   one non-trivial element given in convenient representation by the conjugation $\omega (g) = \omega_1 g \omega_1^{-1}$ with
\begin{equation}\label{eq:OuterSL2}
\omega_1= \begin{pmatrix}
1 & 0 \\
0 & -1
\end{pmatrix}  \not\in SL(2,\mathbb{R}) \ .
\end{equation}
The corresponding Lie algebra automorphism $\Omega$ defined by $\Omega [T_A] = \Omega^B{}_A T_B =  \omega_1 T_A \omega_1^{-1}$ is readily shown to be an involution and therefore defining with \eqref{eq:LambdaGluing} consistent integrable boundary conditions. Note that we might as well represent the non-trivial $\text{Out}_0(SL(2,\mathbb{R}))$ element by a conjugation with,
\begin{equation}\label{eq:OuterSL2v2}
\omega_2 = \begin{pmatrix}
0 & 1 \\ 1 & 0
\end{pmatrix}  \not\in SL(2,\mathbb{R}) ,
\end{equation}
which is connected to $\omega_1$ with an involutive inner automorphism that corresponds in $SO(1,2)$ with a rotation over $\pi$. Moreover, it is important to emphasise  that this inner automorphism is the only allowed group translation leading to integrable D-brane configurations in AdS$_3$. We are thus allowed to make two distinguished cases in the analysis of D-branes in the $\lambda$-deformed AdS$_3$ corresponding to regular and twisted conjugacy classes.

\noindent First, let us parametrise the $SL(2,\mathbb{R})$ in a general way as,
\begin{equation}\label{eq:SL2group}
g = \begin{pmatrix}
X_0 + X_1 & X_2 + X_3 \\
X_2 - X_3 & X_0 - X_1
\end{pmatrix}
\end{equation}
with $X_i$, $i=\{0,1,2,3\}$, real and with the defining relation $-X_0^2 + X_1^2 + X_2^2 - X_3^2 = -1$, making apparent the AdS$_3$ embedding. \\
\indent The regular conjugacy classes (obtained by taking $\omega$ the identity) are distinguished by $\Tr (g) = 2 X_0$ constant and substituting this into the defining relation one finds the geometry of the corresponding D-branes (see also \cite{Bachas:2000fr,Stanciu:1999nx,Figueroa-OFarrill:2000gfl}). The geometry will depend on the values of $X_0$; for $X_0 > 1$ the conjugacy classes correspond to de Sitter D$1$-strings, for $X_0<1$ they correspond to $H_2$ instantons and for $X_0 = 1$ to the future- and past light-cone. In the next, we will consider only  the former de Sitter D$1$-strings since the study of instantons is beyond the goal of this paper and for the latter case the metric will be degenerate. However, as shown by \cite{Bachas:2000fr}, the de Sitter WZW-branes are tachyonic and, as we will shortly touch upon, they will stay so when turning on the deformation.  \\
\indent The twisted  conjugacy classes  \eqref{eq:TwistConjClass} are obtained by  conjugation with $\omega_1$, 
 \begin{equation}\label{eq:AdSTwistConjClass}
 C_\omega (g) = \left\{ h \,g\, \omega_1\, h^{-1} | \; \forall h\in G \right\} \omega_1^{-1},
 \end{equation}
 and are thus distinguished by $\Tr \left(g \, \omega_1 \right) = 2 X_1$ constant. The corresponding D-brane configurations are, for any $X_1\in\mathbb{R}$,  two-dimensional Anti de Sitter D$1$-strings (see also \cite{Bachas:2000fr}). Equivalently, when conjugating with $\omega_2$, twisted conjugacy classes will be distinguished by $\Tr \left(g  \omega_2 \right) = 2 X_2$ constant corresponding again to AdS D$1$-strings. The choice of representation depends on how one wants to analyse these D-brane configurations together with the choice of parametrisation of the $SL(2.\mathbb{R})$ group element. In both cases, however, one sees from the defining AdS$_{3}$ relation that the \textit{integrable} AdS$_2$ branes are  static configurations. In the WZW case, these D-branes configurations are shown to be physical in \cite{Bachas:2000fr}.
 
\noindent {\bf dS D$1$-strings}\\
A convenient parametrisation to describe the dS D$1$-strings is one where we replace in  \eqref{eq:SL2group} the elements by,
\begin{equation}
X_0 = \cosh\psi, \quad X_3 = \sinh\psi \sinh\tau , \quad X_1 + i X_2 = \sinh\psi \cosh\tau e^{i\phi}
\end{equation}
with $\psi \in [0,\infty [$, $\tau \in ]-\infty , \infty[$ and $\phi \in [0,2\pi]$ (although, they are not good global coordinates). Here the AdS$_3$ is build up out of fixed  $\psi$-slices  of dS$_2$ spacetimes parametrised by $\tau$ and $\phi$ corresponding to the dS D$1$-strings. Note that in this coordinate system we describe D$1$-strings that are static configurations in AdS$_3$.

\noindent The target space geometry of the $\lambda$-deformed AdS$_3$ is,
\begin{equation}\label{eq:dSGeometry}
\begin{aligned}
\mathrm{d}s^{2}_{\lambda} &= 2k \left(\frac{1+\lambda}{1-\lambda}\,\mathrm{d}\psi^{2} +\frac{1-\lambda^{2}}{\widehat{\Delta}}\,\sinh^{2}\psi\ \left(- \mathrm{d}t^{2}+ \cosh^{2}t \, \mathrm{d}\phi^{2} \right) \right)  \\
H_\lambda &= 4k \frac{2\lambda \widehat{\Delta} + (1-\lambda^{2})^{2}}{\widehat{\Delta}^{2}}\,\sinh\psi^{2}\cosh\tau \mathrm{d}\psi \wedge \mathrm{d}\tau \wedge \mathrm{d}\phi  \\
e^{-2\Phi} &= e^{-2\Phi_0} \widehat{\Delta}
\end{aligned}
\end{equation}
with $\widehat{\Delta} = 1+\lambda^2 - 2\lambda\cosh 2\psi$. In these coordinates it is obvious that the deformation leaves the dS$_2$ D$1$-strings intact but changing  the radius with a squashing factor $\frac{1-\lambda^{2}}{\widehat{\Delta}}$. 
\indent Comparing the integrable boundary conditions obtained from \eqref{eq:LambdaGluing} with $\Omega = 1$ with the boundary conditions from the sigma model approach, i.e.\ Dirichlet \eqref{eq:dirichlet} and (generalised) Neumann \eqref{eq:neumann}, one can extraxt the two-form $\mathcal{F}$ on the dS D$1$-brane to find,
\begin{equation}
\mathcal{F} = k \frac{(1-\lambda)^{2}}{\widehat{\Delta}} \sinh 2\psi \cosh\tau \mathrm{d}\tau \wedge \mathrm{d}\phi .
\end{equation}
See also \cite{Alekseev:1998mc,Stanciu:2000fz,Gawedzki:1999bq} for a general formula in terms of the gluing matrix. The induced metric $\widehat{G}$ on the D$1$-brane is obtained simply by enforcing $\psi$ = constant in eq.~\eqref{eq:dSGeometry}. The DBI action now evaluates to (absorbing a factor $1-\lambda$ into the constant dilaton),
\begin{equation}
\begin{aligned}
S_{\text{DBI}} =T_1  \int e^{-\Phi} \sqrt{ -\det (\widehat{G}_{ab}+{\cal F}_{ab})} =  4 \pi  k \,T_1 \,e^{-\Phi_0}    \int \mathrm{d}\tau \sqrt{- \sinh^{2}\psi \cosh^{2}\tau }
\end{aligned}
\end{equation}
and, hence, is supercritical.\\
\indent One could equally perform this analysis in the global cylindrical coordinates along the lines of \cite{Bachas:2000fr} to describe dynamical configurations of a circular D$1$-string and argue that these are unphysical trajectories. We expect however no new information to be gained.

\noindent {\bf Anti de Sitter D$1$-strings}\\
At first sight a logical coordinate system to describe the AdS D$1$-strings seems to be the global AdS coordinates where the elements of \eqref{eq:SL2group} are replaced by,
\begin{equation}
X_0 + i X_3 = \cosh\psi \cosh\omega e^{i \tau}, \quad X_1 = \sinh\psi, \quad X_2 = \cosh\psi \sinh\omega
\end{equation}
with $\{\psi, \tau, \omega\} \in \mathbb{R}^3$. The twisted conjugacy class \eqref{eq:AdSTwistConjClass} (obtained by conjugation with $\omega_1$ of eq.~\eqref{eq:OuterSL2}) lie along fixed $\psi$-slices which correspond to AdS$_2$ spacetimes parametrised by $\tau$ and $\omega$.

\newcommand\numberthis{\addtocounter{equation}{1}\tag{\theequation}}

\noindent In this coordinate system  the target space geometry of the $\lambda$-deformed theory is,
\begin{align*}\label{eq:AdSLambda}
\mathrm{d}s^{2}_\lambda &= 2k\, \frac{1+\lambda}{1-\lambda} \left( \frac{\widetilde{\Delta} + 4\lambda(\cos^{2}\tau\cosh^{2}\omega-1)}{\widetilde{\Delta}}\, \mathrm{d}\psi^{2} \,  + 2\lambda \frac{\sin 2\tau \cosh^{2}\omega \sinh 2\psi}{\widetilde{\Delta}} \, \mathrm{d}\psi \mathrm{d}\tau  \right. \\
\qquad\;  & - 2\lambda  \frac{\cos^{2}\tau \sinh 2\omega \sinh 2\psi}{\widetilde{\Delta}}\, \mathrm{d}\psi \mathrm{d}\omega + \cosh^{2}\psi \left(- \frac{(1+\lambda)^{2} -4\lambda\cos^{2}\tau}{\widetilde{\Delta}} \, \cosh^{2}\omega\mathrm{d}\tau^{2} \right. \\
\qquad \; & \left. \left. +2\lambda\frac{\sin 2\tau \sinh 2\omega}{\widetilde{\Delta}}\, \mathrm{d}\tau\mathrm{d}\omega +  \frac{(1-\lambda)^{2}-4\lambda \cos^{2}\tau \sinh^{2}\omega}{\widetilde{\Delta}}\, \mathrm{d}\omega^{2}\right)  \right) \\
H_\lambda &= 4k\, \frac{2\lambda \widetilde{\Delta} + (1-\lambda^{2})^{2}}{\widetilde{\Delta}} \cosh^{2}\psi \cosh\omega \mathrm{d}\psi \wedge \mathrm{d}\tau \wedge \mathrm{d}\omega \numberthis \\
e^{-2\Phi} &= e^{-2\Phi_0} \widetilde{\Delta} 
\end{align*}
with $\widetilde{\Delta} = (1+\lambda^{2}) - 4\lambda \cos^{2}\tau \cosh^{2}\omega \cosh^{2}\psi $. One can readily verify that in the WZW $\lambda\rightarrow 0$ limit the metric reduces to the obvious slicing of AdS$_3$ by AdS$_2$ spacetimes along $\psi$. However, when turning on the deformation this slicing becomes obscure and one can not read of a ``squashing" factor of the AdS D$1$-strings in contrary to the dS case of above. An explicit analysis  of the integrable boundary conditions \eqref{eq:LambdaGluing} with the conjugation by  $\omega_1$ teaches us however that it is indeed the $\psi$ direction that is Dirichlet.  The induced metric $\widehat{G}$ on the brane is thus obtained by enforcing $\psi$ constant in \eqref{eq:AdSLambda}. Comparing again the integrable boundary conditions to the ones obtained from the general sigma model, we can extract the gauge invariant two-form $\mathcal{F}$ on the AdS D$1$-string,
\begin{equation}
\mathcal{F} = k \frac{(1+\lambda)^{2}}{\widetilde{\Delta}} \,\sinh 2\psi \cosh\omega \mathrm{d}\tau \wedge \mathrm{d}\omega .
\end{equation}
The DBI action simply gives,
\begin{equation}
\begin{aligned}
S_{\text{DBI}} =T_1  \int e^{-\Phi} \sqrt{ -\det (\widehat{G}_{ab}+{\cal F}_{ab})} = 2 k\, T_1\, e^{-\Phi_0}    \int \mathrm{d}\tau \mathrm{d}\omega\,  \cosh\psi \cosh\omega ,\end{aligned}
\end{equation}
where we absorbed a factor of $1+\lambda$ into the constant $e^{-\Phi_0}$. The action is readily minimised when $\psi_0 = 0$ where the D$1$-string is still of finite size. However, when making appropriate gauge choices for the induced antisymmetric field $\widehat{B}$ and the $U(1)$ field strength $F$, in particular one where $A_\tau = 0$, one can identify the usual quantisation condition from the Gauss constraint of QED$_2$ (following \cite{Bachas:2000fr}) given by,
   \begin{equation}
  4\pi T_1 e^{-\Phi_0} \frac{ \widetilde{\Delta} \mathcal{F}_{\tau\omega}}{\sqrt{ -\widetilde{\Delta}\, \det (\widehat{G}_{ab}+{\cal F}_{ab})}} = 4\pi T_1 e^{-\Phi_0} \sinh\psi = q \in \mathbb{Z} 
   \end{equation}
with the integer $q$ known to be the number of fundamental  strings bound to the D$1$-string \cite{Witten:1995im}. Similar to the $SU(2)$ example we thus have again, due to a flux quantisation condition, additional locations where the D$1$-strings are stabilised, independent of the value of $\lambda$. However, contrary to the $SU(2)$ case, this does not descend  from topological conditions of the boundary WZW (see also appendix \ref{a:conventions}) as AdS$_3$ is topologically trivial.\\
\indent One could instead also consider the twisted conjugacy class obtained from conjugation by $\omega_2$ of eq.~\eqref{eq:OuterSL2v2}. The corresponding worldvolume is then characterised by $X_2 = \cosh\psi \sinh\omega$ constant and is obtained from the previous fixed $\psi$-slices by a rotation over $\pi$ in the spatial directions. The analysis of these worldvolumes are  easily done in the Poincar\'e coordinates ($t,x,u$) $\in \mathbb{R}^3$ that are obtained by,
\begin{equation}
X_0 + X_1 = u, \ \quad X_0 - X_1 = \left( \frac{1}{u} + u (x+t)(x-t) \right) , \quad X_2 \pm X_3 = u (x\pm t) .
\end{equation}
Eliminating then the coordinate $u$ by $u x = C$ with $C$ a constant one can identify the gauge-invariant two-form ${\cal F}$  to be,
\begin{equation}
\mathcal{F} =  2k \frac{(1+\lambda)^2}{\Delta_P } \, \frac{C^2}{x^2}\, \mathrm{d}t \wedge \mathrm{d} x ,
\end{equation}
with the dilaton factor $\Delta_P$ in Poincar\'e coordinates.  The Gauss constraint similarly quantises the constant $C$ as,
\begin{equation}
4\pi\, T_1 \,e^{-\Phi_0}  C  = q \in \mathbb{Z} ,
\end{equation}
where again a factor of $1+\lambda$ was absorbed in the constant $e^{-\Phi_0}$.

We conclude that from a classical point of view the AdS D$1$-strings in the  $\lambda$ background are still physical. Moreover they are stablised in the same manner  as in the WZW case, i.e. due to flux quantisation.  Semi-classically it would be interesting  to perform a stability analysis of the quadratic fluctuations as  was done for  $SU(2)$ in section \ref{sec:su2}. In the $\lambda$ case we expect  the same stability conclusion as the WZW case (see e.g.\ \cite{Petropoulos:2001qu,Lee:2001xe})   accompanied with a lifting of zero modes. However, we will not pursue this interesting point here.


\section{Relation to generalised T-dualities }\label{sec:Tduals}

\subsection{The non-Abelian T-dual limit} 
In a scaling limit $\lambda\to 1$ the $\lambda$-deformation recovers the non-Abelian T-dual of the principal chiral model \cite{Sfetsos:2013wia}.  To achieve this limit one expands the group element around the identity,
\be
g= \mathbf{1} + \frac{i}{k}v^A T_A +{\cal O} \left( \frac{1}{k^2} \right)  ,
\ee
and takes $k\to\infty$ to find,
\begin{equation}
L_-^A = - \frac{\partial_- v^A}{k} + {\cal O}\left(\frac{1}{k^2}\right), \quad R_+^A = - \frac{\partial_+ v^A}{k} + {\cal O}\left(\frac{1}{k^2}\right), \quad D_{AB} = \eta_{AB} + \frac{F_{AB}{}^{C} v_C}{k} \,.
\end{equation}
In this limit the $\lambda$-deformed action \eqref{eq:LambdaAction2} becomes the non-Abelian T-dual with respect to the $G_L$ action of the PCM \eqref{eq:PCMaction},
\be\label{eq:actNabT}
S = \frac{1}{\pi }\int \partial_+ v_A\left(  {\cal M}^{-1} \right)^{AB} \partial_- v_B  + {\cal O} \left( \frac{1}{k} \right) , \quad \text{with}  \quad  {\cal M} =\kappa^{2} \eta_{AB} + F_{AB}{}^C v_C  \ . 
\ee 
 For the case of $SU(2)$ in  the parametrisation used in eq.~\eqref{eq:su2para} this limit is achieved by taking $\alpha = \frac{r}{ 2k}$  with 
\be
v_1 =- \frac{r}{\sqrt{2}}   \sin \beta \sin \gamma   \,,\quad  v_2 = \frac{r}{\sqrt{2}}  \sin \beta \cos \gamma \, , \quad  v_3 =  \frac{r}{\sqrt{2}} \cos\beta  \ . 
\ee
The metric becomes 
\be
ds^2_{NAbT} = \frac{1}{\kappa^{2}} \left( dr^2 + \frac{r^2 \kappa^{4}}{r^2 +\kappa^{4}} ds^2 (S^2)  \right) \ . 
\ee
where $ds^2 (S^2) = \mathrm{d}\beta^{2}+\sin^{2}\beta \mathrm{d}\gamma^{2}$. The first point to note is that in the limit the two-sphere remains intact so one anticipates that the D-branes described previously are preserved.   Performing the $\lambda\rightarrow1$  limit procedure on the boundary conditions of eq.~\eqref{eq:su2bc} yields,
\be\label{eq:bclim}
\begin{aligned}
\partial_- r &= - \partial_+ r \,,\\ 
(r^2+\kappa^{4}) \partial_-\beta &= (r^2 -\kappa^{4}) \partial_+\beta - 2 r \kappa^{2} \sin \beta\, \partial_+ \gamma\, \\ 
(r^2 +\kappa^{4}) \partial_-\gamma &= 2 r \kappa^{2} \textrm{csc\,} \beta \partial_+ \beta + (r^2 -\kappa^{4}) \partial_+ \gamma \ . 
\end{aligned}
\ee

To understand these conditions it is useful to work instead with the following combination of worldsheet derivatives  
\be
\mathring{L}_+   =  - {\cal M}^{-T} \partial_+ v \ , \quad \mathring{L}_-   =  + {\cal M}^{-1} \partial_- v  \ , 
\ee
which can be used to construct a Lax connection for the dynamics of the non-Abelian T-dual theory  eq.~\eqref{eq:actNabT}: 
\be \label{eq:pcmlax}
\mathring{{\cal L}}_\pm[z] = \frac{1}{1 \mp z} \mathring{L}_\pm \ , \quad [\partial_+ + \mathring{{\cal L}}_+ , \partial_- + \mathring{{\cal L}}_-]= 0  \ . 
\ee
In terms of these the boundary conditions of eq.~\eqref{eq:bclim} take a remarkably simple form
\be
\mathring{L}_+=\mathring{L}_- \ . 
\ee
This property of the boundary conditions holds in general, at least in the case where we set the extra automorphism $\Omega = 1$, and follows due to the limit 
\be
\frac{1}{k+\kappa^2} O_g \to {\cal M}^{-T}  \ ,
\ee
from eq.~\eqref{eq:LambdaGluing}.

Now for the punchline. The  non-Abelian T-dual theory described by eq.~\eqref{eq:actNabT} is classically  equivalent to the principal chiral theory  
\be
S_{\text{PCM}} = -\frac{\kappa^{2}}{\pi} \int \mathrm{d}\sigma\mathrm{d}\tau\ \langle\widetilde{g}^{-1}\partial_+ \widetilde{g}   \widetilde{g}^{-1}\partial_- \widetilde{g}  \rangle \ . 
\ee
The T-duality transformation rules are non-local in terms of the coordinates of the sigma-models but in terms of world sheet derivatives are a canonical transformation of the form 
\be
\widetilde{L}_\pm^a =  \mathring{L}_\pm   \ , \quad  \widetilde{g}^{-1} \partial_\pm \widetilde{g} =-i  \widetilde{L}_\pm^a T_a  \ . 
\ee
Thus we can immediately conclude that under non-Abelian T-duality the  D2-brane described by the boundary condition \eqref{eq:bclim} results in 
\be
\widetilde{g}^{-1} \partial_\sigma \widetilde{g} = 0 \ , 
\ee
i.e. a space-filling D3-brane.

This analysis agrees exactly with the integrable boundary conditions that we would obtain for the principal chiral model by substituting the Lax \eqref{eq:pcmlax} into the  result \eqref{eq:BoundCondLax1}   (note that this holds as the Lax \eqref{eq:pcmlax} satisfies eq.~\eqref{eq:ReflectedTransport}) which leads to the boundary conditions,
\begin{equation}
\mathring{L}_\pm = \Omega (\mathring{L}_\mp) =  \Omega \cdot D^T (\mathring{R}_\mp) \ .
\end{equation}
Interestingly, these have the form of  type N gluing conditions.  In the context of the WZW, these type N gluings eq.~\eqref{eq:typeN}   preserve conformal invariance but break the chiral-algebra (and for which a good understanding is still lacking to our knowledge).  It would be interesting to relate these observations to the recently appeared \cite{Fraser:2018atd}.
 
 \subsection{The pseudo-dual limit }
 A second interesting limit described in \cite{Georgiou:2016iom} is a scaling limit as $\lambda \to -1 $ which results in the pseudo-dual \cite{Nappi:1979ig} of the principal chiral model.   The pseudo-dual theory is obtained by replacing the currents of the PCM with scalars according to 
 \begin{equation}
 \tilde{g}^{-1} \partial^\mu \tilde{g}  = \epsilon^{\mu \nu } \partial_\nu \phi  \ , 
 \end{equation} 
 such that the conservation of the currents becomes a trivial consequence of the commutation of partial derivatives.   The Bianchi identities and equations of motion written in terms of $\phi$ can be obtained from a ``dual'' action.  However this is not a true dualisation  \cite{Nappi:1979ig} -- even at the classical level the two theories are not related by a canonical transformation and at the quantum level they have striking differences.  The PCM is asymptotically free where as the pseudo-dual is not.   The PCM is quantum integrable whereas the pseudo-dual displays particle production.    Nonetheless it is intriguing that the pseudo-dual action follows from the $\lambda$-theory upon the scaling 
 \begin{equation}
 \lambda = -1 + \frac{\kappa^2 }{k^{\frac{1}{3}} }   \ , \quad g =  \mathbf{1} + \frac{i}{k^{\frac{1}{3}} }\phi^A T_A + \dots    ,  \quad k \to \infty \, .
 \end{equation}
Evidently in taking this limit one needs to relax the requirement that $\lambda \in [0,1]$ required of the  original construction of the $\lambda$-model. 

Let us see the effect of this on the boundary conditions in the context of the $SU(2)$ theory.  The scaling is quite similar; we define  $\alpha = \frac{r}{ 2 k^{\frac{1}{3}}}$ and 
\be
\phi^1 =- \frac{r}{\sqrt{2}}   \sin \beta \sin \gamma   \,,\quad  \phi^2 = \frac{r}{\sqrt{2}}  \sin \beta \cos \gamma \, , \quad  \phi^2  =  \frac{r}{\sqrt{2}} \cos\beta  \ . 
\ee
The limit $k\to \infty$ can then be taken in the boundary conditions  of eq.~\eqref{eq:su2bc} resulting simply in  
\be
\partial_\tau \phi^i = 0 \ . 
\ee
In terms of the PCM variables we recover, as with the non-Abelian T limit,  a   D3-brane described by
\be
\widetilde{g}^{-1} \partial_\sigma \widetilde{g} = 0 \ .  
\ee

\subsection{Poisson-Lie dual interpretation}
 
The $\lambda$-deformed theory is closely connected to a class of integrable deformations of the principal chiral model known as $\eta$--deformation (also known as Yang-Baxter deformations) \cite{Klimcik:2008eq,Delduc:2013fga}.    To establish the relation   between the $\lambda$ and $\eta$ theories one first performs an analytic continuation of the coordinates parameterising the $\lambda$-theory and also of the deformation parameter itself.   Performing this analytic continuation  in the $\lambda$-deformed action of \eqref{eq:LambdaAction2} results in a new (real) sigma-model that is the  Poisson-Lie T-dual to the $\eta$-deformation \cite{Hoare:2015gda,Sfetsos:2015nya,Klimcik:2015gba,Hoare:2017ukq}.   Our goal here is to track this connection through with the boundary conditions considered here.  To make this rather technical procedure accessible we first introduce the rudiments  of Poisson-Lie technology. 

 Poisson-Lie T-duality \cite{Klimcik:1995ux} is a generalised notion of T-duality between a pair of $\sigma$-models on group manifolds $\widehat{G}$ and $\widecheck{G}$ (with corresponding algebras $\widehat{\frak{g}}$ and $\widecheck{\frak{g}}$) that do not enjoy isometries   but instead posses a set of currents that are non-commutatively  conserved with respect to the dual algebra  $\widecheck{\frak{g}}$  ($\widehat{\frak{g}}$). 
%
%
For this construction to be consistent $\frak{d} =\widehat{ \frak{g}} \oplus \widecheck{\frak{g}}$ must define a Drinfeld double \cite{Drinfeld:1986in}.   The two Poisson-Lie dual sigma-models defined in this way are of the form, 
\be\label{eq:PLacts}
\begin{aligned}
\widehat{S}[\widehat{g}]&=  \frac{1}{    t \eta} \int d^2 \sigma \widehat{L}_+^T (E_0^{-1}  - \widehat{\Pi})^{-1}\widehat{ L}_- \,,\\ 
\widecheck{S}[\widecheck{g}] &=  \frac{1}{    t \eta} \int d^2 \sigma \widecheck{L}_+^T (E_0   - \widecheck{\Pi})^{-1}\widecheck{L}_- \,, 
\end{aligned}
\ee
in which $\widehat{L}_\pm$($\widecheck{L}_{\pm}$) are pullbacks of left-invariant one-forms for $\widehat{G}$($\widecheck{G}$), $E_0$ is a constant matrix of freely chosen moduli, and $\widehat{\Pi}$($\widecheck{\Pi}$) is a matrix formed by the combination of the adjoint action of $\widehat{G}$($\widecheck{G}$) on itself and $\widecheck{G}$($\widehat{G}$) according to, 
\be
\widehat{g}^{-1}\widehat{ T}_a \widehat{g} = a_a{}^b \widehat{T}_b \ ,  \quad \widehat{g}^{-1} \widecheck{T}^b  \widehat{g} = b^{ab} T_b  + (a^{-1})_b{}^a \widecheck{T}^b \ , \quad \widehat{\Pi}^{ab} = b^{ca} a_{c}{}^b \,,
\ee
where $\widehat{T}_a$ and $\widecheck{T}^a$ resp. are the generators of $\widehat{\frak{g}}$ and $\widecheck{\frak{g}}$ resp. 
The overall tension of the sigma-models has been introduced for later convenience.   What will be useful in our consideration is that the two PL models are canonically equivalent \cite{Sfetsos:1996xj,Sfetsos:1997pi} with the canonical transformation defined by,
 \be
 \begin{aligned}\label{eq:PLCT}
\widehat{ P} = -\widecheck{\Pi} \widecheck{P} + \frac{t\eta}{2}  \widecheck{L}_\sigma \ , \quad  \widecheck{P} = - \widehat{\Pi} \widehat{P} +  \frac{t\eta}{2} \widehat{L}_\sigma \ , 
 \end{aligned}
 \ee
 in which, if we let $X^\mu$ be local coordinates on $\widehat{G}$, we define the momentum $\widehat{P}_a  = \widehat{L}_a^\mu \frac{\delta S}{\delta \dot X^\mu} $\footnote{We adapt the results of the \cite{Sfetsos:2013wia} to our conventions and restore the overall normalisation of the sigma models.}.  
 
 In the context of the $\lambda$--$\eta$ connection the relevant Drinfeld double is $\frak{d} = \frak{g}^{\mathbb{C}}$ and $\widecheck{\frak{g}}$ is identified with a Borel sub-algebra coming from the Iwasawa decomposition $ \frak{g}^{\mathbb{C} } = \frak{g} +  \frak{a} + \frak{n}$.   The action $\widecheck{S}$ is defined on the group manifold $\widecheck{G} \cong AN$ and is the one obtained by analytic continuation of the $\lambda$--deformation.   The dual action, i.e. the first of  eq.~\eqref{eq:PLacts}, is defined on the group manifold $G$ and can be recast as, 
 \be\label{eq:etaact}
 \widehat{S}=  \frac{1}{ t  } \int d^2 \sigma \widehat{R}_+^T (\mathbf{1} - \eta  {\cal R} )^{-1} \widehat{R}_- \, ,
 \ee
 in which $\widehat{R}_\pm$ are right-invariant one forms (pulled back) and ${\cal R}$ solves the modified classical Yang-Baxter equation.   For the isotropic single parameter deformations considered here $E_0^{-1} = \frac{1}{\eta} \bf{1} + {\cal R} $.  This is the integrable $\eta$-deformation \cite{Klimcik:2008eq,Delduc:2013fga}.

  For didactic purpose we consider the case of the $SU(2)$ $\lambda$-deformation. In the parametrisation used in eq.~\eqref{eq:su2para} this analytic continuation amounts to a mapping between coordinates   $(\alpha,\beta, \gamma) \to ( y_1, y_2, \chi)$ and  parameters $(k, \lambda) \to ( t, \eta)$ given by,  
\begin{equation}\label{eq:ana}
y_1 + i y_2  =  i \sin \alpha \sin \beta   e^{i \gamma} \ ,  \quad e^\chi = \cos \alpha +  i \sin \alpha \cos\beta  \ , \quad k = \frac{i}{4 t \eta} \ , \quad \lambda = \frac{i -\eta}{i+\eta}  \ .
\end{equation}
In this case the $\widehat{\frak{g}} = \frak{su}(2)$ and $\widecheck{\frak{g}} = \frak{e}_3$ (Bianchi II) and we will work with   group elements   parametrised by, 
\be
 \widecheck{g} = \left( \begin{array}{cc}  e^{\frac{\chi}{2} } &  e^{-\frac{\chi}{2} }(y_1 - i y_2) \\ 0 & e^{-\frac{\chi}{2} }  \end{array} \right) \ , \quad \widehat{g}=\left(
\begin{array}{cc}
 e^{\frac{1}{2} i (\phi +\psi )} \cos \left(\frac{\theta }{2}\right) & e^{\frac{1}{2} i (\phi -\psi )} \sin \left(\frac{\theta }{2}\right) \\
 -e^{-\frac{1}{2} i (\phi -\psi )} \sin \left(\frac{\theta }{2}\right) & e^{-\frac{1}{2} i (\phi +\psi )} \cos \left(\frac{\theta }{2}\right) \\
\end{array}
\right) . 
\ee
Applying the analytic continuation eq.~\eqref{eq:ana} to the boundary conditions eq.~\eqref{eq:su2bc} yields a result that is real (as required to be a consistent boundary condition)    and rather elegant  when written in terms of the momentum $\widecheck{P}$:  
\be
\begin{aligned}\label{eq:E3bc} 
\widecheck{P}_1 &= \frac{4}{t \eta} \frac{ e^{2 \chi} } {1+ e^{2 \chi} + r^2}  \widecheck{L}_{2\sigma} \ , \\ 
\widecheck{P}_2 &= -\frac{4}{t \eta} \frac{ e^{2 \chi} } {1+ e^{2 \chi} + r^2}  \widecheck{L}_{1 \sigma} \ , \\
\widecheck{P}_3 &= 0  \ . 
\end{aligned}
\ee
Notice that in these conditions all the complicated dependence on the deformation parameter $\eta$  (notwithstanding the factors of $t\eta$) is subsumed into the momenta $\widecheck{P}$.   In this form we can now immediately  apply the canonical transformation of eq.~\eqref{eq:PLCT}  to deduce the corresponding boundary condition for the $\eta$-deformed theory.  Actually something rather special happens;  the canoncial transformation  eq.~\eqref{eq:PLCT}  depends explicitly  not only on  the momenta $\widecheck{P}, \widehat{P}$ and the left-invariant forms $\widecheck{L}_\sigma, \widehat{L}_\sigma$ but also on  all the coordinates through the matrices $\widecheck{\Pi}$ and $\widehat{\Pi}$ in a rather complicated fashion.    It is then by no means  guaranteed that when the  canonical transformation is applied to the boundary conditions of eq.~\eqref{eq:E3bc} that what results will depend only on the coordinates $(\theta, \phi, \psi)$  that parametrise  $\widehat{G}$.  Reassuringly, however, this  does indeed transpire to be the case! 

We are now in a position to present the boundary conditions of the $\eta$-deformed principal chiral model obtained in this fashion.  In terms of the coordinates themselves the boundary condition takes a rather simple form,
\be
\begin{aligned}\label{eq:etabc2}
 \partial_\sigma \psi + \sec \theta \partial_\sigma \phi &= 0 \, ,  \\   
  \eta \partial_\tau \theta + \tan \theta \partial_\sigma \phi &= 0  \, , \\ 
  \eta \partial_\tau \psi - \sec\theta \partial_\sigma \theta & = 0  \, . 
\end{aligned} 
\ee
For reference the geometry corresponding to the $\eta$-deformed theory reads 
\be
\begin{aligned} 
ds^2  &= \frac{1}{t }  \left(  \left( d\phi +\cos \theta d\psi \right)^2 + \frac{1}{1+\eta^2} \left(d\theta^2 +\sin^2 \theta d\psi^2  \right)  \right)\,, \\
B &= \frac{   \eta }{t (1+\eta^2) }   \sin \theta d\theta \wedge d\psi\, , \quad 
H = dB=  0 \,.
\end{aligned} 
\ee 
Since all the coordinates enjoy (generalised) Neumann boundary condition we are describing here a space-filling brane supported by a worldvolume two-form  ${\cal F} = B + 2\pi \alpha^\prime F$.  Making use of the above metric we can readily extract this two-form, 
\be
{\cal F} = \frac{\eta}{t (1+\eta^2)} \sin \theta d\theta \wedge d\psi  \, ,
\ee
showing that $F=dA= 0$. 

 It is also illuminating to express the   results in terms of right invariant forms , 
\be
\widehat{R}^1 = - \cos\phi \sin\theta d\psi + \sin\phi d\theta \ , \quad \widehat{R}^2 = \sin \phi \sin\theta d\psi +\cos\phi d\theta \ ,\quad \widehat{R}^3 = d\phi +\cos \theta d\psi   \, . 
\ee
such that the boundary conditions take the conventional form of a gluing, 
\be 
\widehat{R}_+^i = \mathbb{R}^i{}_j  \widehat{R}_-^j \, , 
\ee
with
\be
 \mathbb{R} =  \mathbb{O}_+^{-1} \mathbb{O}_-  \, , \quad \mathbb{O}_\pm = \frac{1 }{ 1\pm \eta {\cal R}  }   \, .
\ee 
 It is easily verified that $\mathbb{R}$ so defined is an algebra automorphism. 
It is worth emphasising that here the gluing  between currents after the generalised duality is again with an overall plus sign (it is of the form of a WZW N-type boundary condition eq.~\eqref{eq:typeN}) whereas in the original $\lambda$--deformed WZW the gluing  between currents was with an overall minus sign (i.e.\ of WZW D-type eq.~\eqref{eq:WZWgluing}).

To close the circle we can again relate these boundary conditions to the general integrable boundary condition construction.  First we recall that the Lax for the $\eta$-deformed PCM eq.~\eqref{eq:etaact} is given by \cite{Klimcik:2008eq,Delduc:2013fga},
\begin{equation}
\mathcal{L}_\pm( \eta ,z) = \frac{1+\eta^2}{ 1 \pm z }\, \textrm{ad}_{\widehat{g}^{-1}} \cdot \mathbb{O}_\pm\cdot \widehat{R}_\pm  \ .  
\end{equation} 
Using this we can readily see that boundary condition above is obtained from 
\begin{equation}
\mathcal{L}_\tau(\eta, z) = \mathcal{L}_\tau(\eta, - z) 
\end{equation}
and hence is of the form of  the integrable boundary condition that one would obtain from eq.~\eqref{eq:BoundMon1} in which the extra automorphism $\Omega = 1$ and when would choose the freedom to change the deformation parameter conveniently (see also the discussion at the end of section \ref{sec:intbcmethod}).     Actually there is a second possibility, 
\begin{equation}
\mathcal{L}_\tau(\eta, z) = \mathcal{L}_\tau(-\eta, - z) 
\end{equation}
which also gives an integrable condition; this is just $\widehat{R}_+ = \widehat{R}_-$\footnote{ This later choice however appears incompatible with PL T-duality, this is easily seen since the boundary condition is equivalent to $\widehat{L}_\sigma = 0$ and making use of the canonical transformation eq.~\eqref{eq:PLCT} this leads to a PL dual condition $0= \frac{t \eta}{2} (1- \widehat{\Pi} \widecheck{\Pi}) \widecheck{P} + \widehat{\Pi} \widecheck{L}_\sigma$ which still depends on the coordinates of $\widehat{g}$ and is thus non-local.}.


 

\section{Conclusions}

We have seen that integrable boundary conditions of the $\lambda$-deformed theory can be obtained by demanding  that the monodromy matrix  of the Lax connection generates conserved charges even in the presence of a boundary.  Rather elegantly these boundary conditions can be described by (twisted) conjugacy classes, independent of the deformation parameter and indeed as the deformation is turned off the known D-brane configurations in WZW models are recovered.   For the $SU(2)$ theory the picture is nice; viewing $S^3$ as a two-sphere fibred over an interval the conjugacy classes correspond to  D2-branes  wrapping the two-sphere  (that shrink at the end points of the interval to D0-branes), and the effect of the $\lambda$-deformation is in essence to determine the size of the two-spheres.  The quantisation of the world-volume flux remains consistent in the deformed theory---all occurrences of the deformation parameter cancel---and enforce that the D-branes sit at localised positions along the interval. 
\\
\\
Armed with the integrable D-branes of the $\lambda$-model we were then able to show their connection to D-branes in the PCM and its $\eta$-deformation.  First we could track the D-brane boundary condition through to the non-Abelian T-dual point ($\lambda =1$) and dualise them to an N-type boundary condition of the PCM, which is also integrable.   Alternatively we could perform analytical continuation to ascertain boundary conditions for a Poisson-Lie sigma model on the group manifold $AN$ appearing in the Iwasawa decomposition $G^{\mathbb{C}}= G A N$.  The boundary conditions produced in this fashion were then Poisson-Lie T-dualisable and produced  N-type boundary conditions of the $\eta$-deformed PCM.  Again we saw explicitly that these D-branes of the $\eta$-deformed PCM are integrable.   The exchange of $N$-type and $D$-type boundary conditions in this approach is a phenomenon that seems generic   in the context of non-Abelian and Poisson-Lie theories. 
\\
\\
Let us comment on a few interesting open problems triggered by this study. 
\\
\\
The concerns and analysis of this paper have been predominantly classical.  An important next direction is to make more precise the quantum description corresponding to the boundary conditions considered here.   Assuming no Goldschmidt-Witten anomaly is encountered one might anticipate that the integrability transfers to the quantum theory.  Here the situation is rather intriguing.   A bulk S-matrix is conjectured -- and to a certain extent derived via quantum inverse scattering -- for the $\lambda$-deformation and has a factorised product form $S(\theta) = X(\theta) S^{SU(2)}(\theta) \otimes S^{RSOS_k }(\theta)$ (for the $SU(2)$ theory),  in which the first factor is the $SU(2)$ rational solution of the Yang-Baxter equation and the second is a interaction round a face type block that is thought of as describing kink degrees of freedom \cite{Appadu:2017fff}.   A quantum integrable boundary should supplement this bulk S-matrix with a boundary `K-matrix' that obeys a boundary version of the Yang-Baxter equation  \cite{Cherednik:1985vs,Sklyanin:1988yz} .   It will be interesting, and the subject of further investigation, to establish the boundary K-matrix corresponding to integrable boundary conditions found within.     Developing this line further, the quantum inverse scattering construction shows how the $\lambda$-theory can be quantised on a lattice as a spin $k$ XXX Heisenberg chain with impurities \cite{Appadu:2017fff}.    It is appealing to establish a match between integrable boundary conditions of such spin chains (studied e.g.\ in \cite{deVega:1993xi}) and the boundary conditions of the continuum theory we constructed here. 
\\
\\
Here we have considered just bosonic  $\lambda$-theory on a group manifold. These   $\lambda$-deformations have an analogue in the context of symmetric spaces \cite{Hollowood:2014rla}   (i.e. deformations of gauged WZW models) which will be of interest to study, with the anticipation that the   geometric description of D-branes of \cite{Maldacena:2001ky,Gawedzki:2001ye,Fredenhagen:2001kw,Stanciu:1997sk} persists in the deformed theory.  Going further one can consider $\lambda$-deformations of theories based on supercosets with applications to the $AdS_5 \times S^5$ superstring  \cite{Hollowood:2014qma,Appadu:2015nfa}.  Here the deformation is expected to be truly marginal and conjectured to correspond to a root-of-unity deformation of the holographic dual gauge theory.  The study of the integrable D-branes in this arena also seems profitable.
\\
\\
One way to introduce fermionic degrees of freedom is by considering supergroups or supercosets as target manifolds as outlined in the previous paragraph. Another way is through the supersymmetrization of the deformed $\sigma $-model thereby introducing worldsheet fermions. We expect that the results obtained in this paper carry over unchanged to the $N=(1,1)$ supersymmetric version of the isotropically $\lambda $-deformed theory. However as is well known, going to $N=(2,1)$ or $N=(2,2)$ supersymmetry, which is needed when one has string theoretical applications in mind, requires additional geometrical structure(s) thereby strongly restricting the allowed target manifolds and the choices of metric and torsion on it. E.g., on integrable Yang-Baxter deformation of the PCM with a Wess-Zumino term (which is a generalization of the $\eta$-deformed PCM) it is highly unlikely that one can go beyond $N=(1,1)$ supersymmetry \cite{Demulder:2017zhz}. As far as we know, the question whether $\lambda $-deformed theories allow for extended supersymmetry, even in the absence of boundaries, has not been addressed yet and forms an interesting open question.

\section*{Acknowledgments}

\noindent
DCT is supported by a Royal Society University Research Fellowship {\em Generalised Dualities in String Theory and Holography} URF 150185 and in part by STFC grant ST/P00055X/1. This work is supported in part by the ``FWO-Vlaanderen'' through the project G020714N and one ``aspirant'' fellowship (SD), and by the Vrije Universiteit Brussel through the Strategic Research Program ``High-Energy Physics''.    We thank Saskia Demulder, Tam\'as Gombor, Tim Hollowood, Marios Petropoulos and Kostas Sfetsos for useful conversations/communications that aided this project. 
  \appendix


\section{General sigma models with boundaries}\label{a:sigmamodel}

To establish our sigma model conventions we briefly review the necessary basics of bosonic open strings in general curved backgrounds (see for instance \cite{Schomerus:2002dc}). We discard the dilaton in this brief discussion and adapt throughout this paper the open string picture. 

The string sigma model is a theory of maps $X^{\mu}(\tau, \sigma)$ from the worldsheet $\Sigma$ parametrised by $(\tau,\sigma)$  to a target space manifold $\mathcal{M}$ parametrised by $X^\mu$ with $\mu\in\{0,\cdots, D-1\}$. Considering open strings, the worldsheet $\Sigma$ has a boundary $\partial\Sigma$ that is mapped in the target space to a $p+1$-dimensional submanifold\footnote{We do not consider the possibility of intersecting D-branes nor a stack of D-branes here.} $N \subset \mathcal{M}$ known as a D$p$-brane. For a target space with metric  $G_{\mu\nu}(X)$ and anti-symmetric 2-form  $B_{\mu\nu}(X)$ the   sigma model action is 
\begin{align}
S_{\sigma} = \frac{1}{4\pi\alpha'} \int \mathrm{d}^{2}\sigma \sqrt{-g} \partial_\alpha X^\mu \left( g^{\alpha\beta} G_{\mu\nu}(X)   + \epsilon^{\alpha\beta} B_{\mu\nu} (X) \right)  \partial_\beta X^\nu  +  \int_{\partial \Sigma} \mathrm{d}\tau A_a (X) \frac{\mathrm{d} X^a}{\mathrm{d}\tau} \ ,
\end{align}
with  $A_a (X)$ the $U(1)$ gauge field coupling to the end-points of the open string and $a \in \{0,\cdots p\}$ denoting the tangent directions to the worldvolume of the D$p$-brane. In conformal gauge, $g_{\alpha\beta} = \text{diag}(+1 , -1)$, and lightcone coordinates, $\sigma^{\pm} = \tau \pm \sigma$, we have
\begin{equation}\label{eq:GenActionLightCone}
S_{\sigma} = \frac{1}{\pi\alpha'}\int\mathrm{d}\sigma\mathrm{d}\tau\, \partial_+ X^\mu \left( G_{\mu\nu}(X) + B_{\mu\nu}(X) \right) \partial_- X^\nu + \int_{\partial\Sigma}\mathrm{d}\tau\, A_a (X) \frac{\mathrm{d} X^a}{\mathrm{d}\tau}  .
\end{equation}
Varying the action with respect to the fields $X^\mu$ (to obtain the equations of motion) one  encounters a boundary term leading to Dirichlet and (generalised) Neumann directions provided the metric splits orthogonally:
\begin{alignat}{3}
&\text{Dirichlet:} \quad &&\delta X^{\widehat{a}} |_{\partial\Sigma} = 0\quad \Rightarrow\quad \partial_\tau X^{\widehat{a}}|_{\partial\Sigma} = 0 \ ,\label{eq:dirichlet}\\
&\text{Neumann:} \quad && G_{ab}(X) \partial_\sigma X^b |_{\partial\Sigma} =\left( B_{ab}(X) + 2\pi \alpha' F_{ab}(X) \right) \partial_\tau X^b|_{\partial\Sigma} \ ,  \label{eq:neumann}
\end{alignat}
with $\widehat{a}=p+1,\ldots, D-1$ the directions normal to the D$p$-brane and where we introduced the Abelian field strength $F_{ab}(X) = \partial_a A_b(X) - \partial_b A_a(X)$. The classical energy momentum tensor of the sigma model is given by
\begin{align}\label{eq:EMGen}
T_{\alpha\beta}
&  = \partial_\alpha X^\mu G_{\mu\nu}(X) \partial_\beta X^\nu - \frac{1}{2} g_{\alpha\beta} g^{\gamma\delta} \partial_\gamma X^\mu G_{\mu\nu}(X) \partial_\delta X^\nu \ .
\end{align}
and in light cone coordinates we have,
\begin{equation}
T_{\pm\pm} = \partial_\pm X^\mu G_{\mu\nu}(X) \partial_\pm X^\nu, \qquad T_{+-} = 0 \ .
\end{equation}
It is straightforward to see that on the boundary, imposing either Dirichlet conditions \eqref{eq:dirichlet} or generalised Neumann conditions \eqref{eq:neumann}, the energy-momentum tensor satisfies
\begin{equation}\label{eq:classicalconfcond}
T_{++}|_{\partial\Sigma} = T_{--}|_{\partial\Sigma} \quad  \rightarrow \quad T_{10}|_{\partial\Sigma} = 0 \ .
\end{equation}
which we will call the (classical) conformal boundary condition. Hence,  there is no momentum flow through the boundary (although $A$ and $B$ charge can be interchanged). If we now summarise the Dirichlet and Neumann conditions (\ref{eq:dirichlet},~\ref{eq:neumann}) using a map $W$ that combines them as,
\begin{equation}
\partial_+ X^{\mu} \vert_{\partial_\sigma}=\left. W^{\mu}{}_\nu \partial_- X^\nu \right\vert_{\partial_\sigma} ,
\end{equation}
then the Dirichlet conditions  correspond to $-1$ eigenvalues of $W$ while the generalised Neumann conditions are described by all other eigenvalues. The classical conformal boundary condition eq.~\eqref{eq:classicalconfcond} then requires that  $W$ preserves the target space metric $G$, 
\begin{equation}\label{eq:SigmaModelBC}
W^T G W = G \ .
\end{equation}

\noindent The dynamics of the D$p$-brane with tension $T_p$ is governed by the DBI action (throughout this paper we ignore the scalar fields parameterizing the fluctuations transversal to the brane),
\begin{equation}\label{eq:DBIaction}
S_{\text{DBI}} = T_p \int  e^{-\Phi} \sqrt{\det(\widehat{G}_{ab}(X) +  \mathcal{F}_{ab}(X))} ,
\end{equation}
where $\widehat{G}_{ab}(X)$ is the induced metric on the worldvolume and $\mathcal{F}_{ab}(X) $ is the gauge-invariant worldvolume flux given by 
\begin{equation}\label{eq:worldvolumeflux}
\mathcal{F}_{ab}(X)  = \widehat{B}_{ab}(X) + 2\pi\alpha' F_{ab}(X) \ ,
\end{equation}
with $ \widehat{B}_{ab}(X) $ the induced anti-symmetric 2-form.


%
%

\section{WZW models and conventions}\label{a:conventions}

In this appendix we collect a number of conventions together with a short review concerning (boundary) WZW models. The WZW model is a non-linear sigma model of maps $g(\tau,\sigma)$ from a 1+1 dimensional Riemann surface $\Sigma$ (with or without boundary) to a Lie group $G$. The model is exact conformal invariant and hence simple enough to describe strings propagating in curved backgrounds. 

Before writing down the action let us make our conventions clear. We pick for the Lie algebra a basis of hermitian generators  $\{T_A \}$, with $A = 1, \ldots , \text{Dim}(G)$, that satisfy $\left[ T_A , T_B \right] = i F_{AB}{}^C T_C$.  The ad-invariant metric on the Lie algebra  is given by $\langle T_A , T_B \rangle =\frac{1}{x_r} \Tr (T_A T_B ) = \eta_{AB}$ with $x_r$ the index of the representation $r$. The left and right-invariant Maurer-Cartan one-forms are expanded in the Lie algebra as $L = g^{-1}\mathrm{d}g = - i L^A T_A$ and $R = \mathrm{d}g g^{-1} = -i R^A T_A$ respectively. They are related by an adjoint action $D(g)\left[T_A\right] = D^B{}_A(g) T_B = g T_A g ^{-1}$ so that $D_{AB}(g) = \Tr \left( g^{-1} T_A g T_B\right)$ and  $D^T_{AB} (g) = D_{AB}(g^{-1})$. Hence, $R^A = D^A{}_{B}(g) L^B$. 

\noindent The  WZW action \cite{Witten:1983ar} is 
 \be \label{eq:WZWaction1}
  S_{\text{WZW,k}}(g) = -\frac{k}{2\pi}\int_\Sigma  d  \sigma d\tau    \langle g^{-1} \partial_+ g , g^{-1} \partial_- g \rangle - \frac{  k}{4\pi }       \int_{M_3}  H   \ , 
   \ee
   where $H$ is the closed torsion 3-form (locally satisfying $H = \mathrm{d}B$) given by
   \begin{equation}
 H = \frac{1}{6}   \langle   \bar g^{-1} d\bar g, [\bar g^{-1} d\bar g,\bar g^{-1} d\bar g]  \rangle ,
   \end{equation}
 with $\bar{g}$ the extension of $g$ into $M_3 \subset G$ such that $\partial M_3 = g(\Sigma)$. There are two  topological obstructions for the consistency of the definition of the WZW action and its quantum theory\footnote{By  construction it is obvious that these obstructions still apply for the $\lambda$ model \eqref{eq:LambdaAction1}.}. First, the existence of $M_3$ is guaranteed only when the second homology group $H_2(G)$ is empty. Second, the path integral based on this action is insensitive to the choice of extension provided that the third cohomology class $[H]/2\pi\in H^3(G)$ is integral. For $SU(2)$ we  have $ H^3(SU(2))\cong \mathbb{Z}$ requiring  the level $k$ to be integer quantised while on the other hand for $SL(2,\mathbb{R})$ we have $H^3(SL(2,\mathbb{R}))$ empty which allows the level $k$ to be free. We want to emphasise here that comparing the WZW action, which  in terms of vielbeins is,
  \be
   S_{\text{WZW,k}}(g) = \frac{k}{2\pi}\int_\Sigma  d  \sigma d\tau   L^A_+ \eta_{AB} L_-^B    
   +  \frac{  k}{24\pi }       \int_{M_3}  F_{ABC} L^A \wedge L^B \wedge L^C ,
   \ee
   to a worldsheet model \eqref{eq:GenActionLightCone} we have units in which $\alpha' = 2$, crucial for $k\in\mathbb{Z}$ when $G=SU(2)$.\\
\indent The WZW model is invariant under a global $G(z) \times G(\bar{z})$ action leading to an infinite-dimensional  symmetry group described by the chirally conserved  holomorphic Kac-Moody currents\footnote{Note that whether the left current is holomorphic or anti-holomorphic depends on the sign of the WZ term.}
 \begin{equation}\label{eq:WZWcurrents}
   J(z) = -k \,\partial g g^{-1}, \qquad \bar{J}(\bar{z}) = k\,  g^{-1}\bar{\partial}g \ ,
   \end{equation}
in the conventions $z = x^0 + i x^1 = i \sigma^+$ and $\bar{z}=x^0 - i x^1 = i \sigma^-$  with the Euclidean worldsheet coordinates $(x^0,x^1) = (i \tau, \sigma)$ and lightcone coordinates $\sigma^{\pm} = \tau \pm \sigma$. At the quantum level the current algebra takes the form,
\begin{equation}\label{eq:WZWCurrentAlgebra}
J^A(z)J^B(w) = \frac{i F^{AB}{}_C J^C (w)}{z-w} + \frac{k \eta^{AB}}{(z-w)^{2}} + \text{reg.}\,,
\end{equation}
and analogous for the $\bar{J}\bar{J}$ OPE (hence the sign difference in the definition \eqref{eq:WZWcurrents}). The exact conformal invariance is established through  the energy-momentum tensor obtained via the Sugawara construction based on the current algebra \cite{DiFrancesco:1997nk},
\begin{equation}
T(z) = T_{zz}(z)= \frac{1}{2(k+h^\vee)} \eta_{AB}  (J^A J^B) (z) ,
\end{equation}
where $h^\vee$ is the dual Coxeter number of $G$ (here we assumed for simplicity  $G$ to be semi-simple). The central charge of the theory can then be found to be,
\begin{equation}
c = \frac{k \,\text{dim}(G)}{k+h^\vee }\ ,
\end{equation}
and all analogous for $\bar{T}(\bar{z})$.

 
 When considering a boundary in the  WZW model one will seek boundary conditions that preserve its exact conformal invariance. Since the CFT is easily described in terms of the chiral currents \eqref{eq:WZWcurrents} it is convenient to express these boundary conditions as a class of gluing conditions\footnote{To compare these gluing conditions (which take value in $T_e G$) to sigma model boundary conditions of the form  \eqref{eq:SigmaModelBC} one should still  translate them to $T_g G$.} for the currents at $z = \bar{z}$,
 \begin{equation}\label{eq:WZWgluing}
 J(z)| = \Omega \bar{J}(\bar{z}) | ,
 \end{equation}
 with $\Omega : \mathfrak{g} \rightarrow \mathfrak{g}$. To preserve the exact conformal invariance the gluing condition should  satisfy the conformal boundary  condition $T(z)| = \bar{T}(\bar{z})|$ . This translates into the condition that the gluing map $\Omega$ should be an isometry of the ad-invariant Lie algebra metric. One could further require $\Omega$ to be an algebra automorphism and, hence, the gluing condition to preserve also the infinite-dimensional current algebra \eqref{eq:WZWCurrentAlgebra}.  The corresponding D-brane configurations are  well-understood and known as symmetric D-branes \cite{Kato:1996nu,Stanciu:1999id,Alekseev:1998mc,Felder:1999ka} or of `type D'  \cite{Stanciu:1999id}, a terminology that we will adapt. Another possibility analysed in \cite{Stanciu:1999id,Stanciu:1999nx}, where it was dubbed `type N', is to consider,
\begin{equation}\label{eq:typeN}
  J(z)| = - \Omega \bar{J}(\bar{z}) |.
 \end{equation} 
 with $\Omega$ a metric-preserving automorphism. They  preserve the conformal invariance but do not preserve the current algebra which makes them somewhat more difficult to analyse. However, as  suggested in section \ref{sec:Tduals} they seem to be related  by generalised T-dualities in the context of the $\lambda$ deformation of the WZW model.\\
\indent  From the sigma model point of view the WZW action \eqref{eq:WZWaction1} is  necessarily modified when the Riemann surface $\Sigma$ has a boundary $\partial\Sigma$ \cite{Klimcik:1996hp,Gawedzki:1999bq,Figueroa-OFarrill:2000lcd,Stanciu:2000fz}.  The image of $g(\partial\Sigma)$  is a (D-brane) submanifold $N$ of $G$  on which a two-form $\omega$ lives such that the restriction of $H$ on $N$ coincides with $\mathrm{d}\omega$. Locally the two-form coincides with the gauge-invariant worldvolume flux $\mathcal{F}$, i.e.\ $\omega =\widehat{B} +4\pi\mathrm{d}A$ \cite{Figueroa-OFarrill:2000lcd}. The  action of the boundary WZW model is,
\begin{equation}\label{eq:boundaryWZW}
S_{\text{WZW},k}(g) =  -\frac{k}{2\pi}\int_\Sigma  d  \sigma d\tau    \langle g^{-1} \partial_+ g , g^{-1} \partial_- g \rangle - \frac{  k}{4\pi }       \int_{M_3}  H + \frac{k}{4\pi} \int_{D_2} \omega   \ , 
\end{equation}
where $M_3 \subset G$ with boundary $\partial M_3 = g(\Sigma) + D_2$ and $D_2 \subset N$. Note that only the boundary equations of motion  will depend on the two-form $\omega$. Demanding that the boundary conditions obtained from the gluing conditions that preserve conformal invariance  \eqref{eq:WZWgluing} coincides with the boundary conditions from the sigma model approach \eqref{eq:dirichlet} and \eqref{eq:neumann} will completely determine the two-form $\omega$ on the D-brane in terms of the gluing map $\Omega$ as in  \cite{Alekseev:1998mc,Stanciu:2000fz,Gawedzki:1999bq}.  \\
\indent Again, there are two  topological obstructions for the consistency of the definition of the boundary WZW action and its quantum theory (for a detailed exposition see \cite{Klimcik:1996hp,Gawedzki:1999bq,Figueroa-OFarrill:2000lcd}) . The existence of $M_3$ and $D_2$ is guaranteed only when the second relative homology $H_2 (G,N)$ vanishes. The path integral is insensitive to the choice of $M_3$ and $\omega$ provided that the third relative cohomology class $\left[\left(H,\omega\right)\right] \in H^3(G,N)$ is integral. As seen in section \ref{sec:examples} this condition enforces for $G = SU(2)$ the position of D2-branes to sit on only a discrete number of values.  Locally this quantisation condition coincides with the quantisation of the worldvolume flux of the $U(1)$ gauge-field $A$ on the brane  \cite{Figueroa-OFarrill:2000lcd,Bachas:2000ik}. For   $G=SL(2,\mathbb{R})$ the position of the D$1$-strings will not be constrained by this particular topological obstruction; however,  the D$1$-strings carry a natural quantisation descending  from the Gauss constraint of two-dimensional gauge theory on the brane \cite{Bachas:2000fr,Witten:1995im}.

\end{document}